\pdfoutput=1
\documentclass[a4paper,11pt]{article}
\usepackage{graphicx}
\usepackage{caption}
\usepackage{subcaption}
\usepackage{jheppub-mod}

\usepackage[utf8]{inputenc}
\usepackage[T1]{fontenc}
\usepackage{amsmath}
\usepackage{amsfonts}
\usepackage{amssymb}
\usepackage{braket}

\graphicspath{{plots/}}

\newcommand{\rev}[1]{#1}

\newcommand{\ms}{\mskip 1.5mu}

\newcommand{\half}{\tfrac{1}{2}}
\newcommand{\dd}{\mathrm{d}}
\newcommand{\Op}{\mathcal{O}}
\newcommand{\tr}{\operatorname{tr}}

\newcommand{\fm}{\operatorname{fm}}
\newcommand{\mev}{\operatorname{MeV}}
\newcommand{\gev}{\operatorname{GeV}}

\newcommand{\tvec}[1]{\boldsymbol{#1}}
\newcommand{\mvec}[1]{\vec{\mskip 0.5mu #1}\mskip 1.5mu}

\newcommand{\lsim}{\;\scalebox{0.85}{$\lesssim$}\;}

\title{Extracting Mellin moments of \\
   double parton distributions from lattice data}

\author[a]{Markus Diehl,}
\author[a,b]{Oskar Grocholski,}
\author[c]{Daniel Reitinger,}
\author[c,d]{Andreas Sch{\"a}fer}
\author[e,f]{and Christian Zimmermann}

\affiliation[a]{Deutsches Elektronen-Synchrotron DESY, Notkestr~85, 22607
   Hamburg, Germany}
\affiliation[b]{Irfu, CEA, Universit{\'e} Paris-Saclay, F-91191, Gif-sur-Yvette,
   France}
\affiliation[c]{Institute for Theoretical Physics, University of Regensburg,
   93040 Regensburg, Germany}
\affiliation[d]{Department of Physics, National Taiwan University, Taipei 106,
   Taiwan}
\affiliation[e]{Department of Physics and Astronomy, University of Kentucky,
   Lexington, KY 40506, USA}
\affiliation[f]{Nuclear Science Division, Lawrence Berkeley National Laboratory,
   Berkeley, CA 94720, USA}

\abstract{Reconstructing Mellin moments of double parton distributions from
calculations on a Euclidean lattice requires taking an integral over a variable
that may be regarded as a Ioffe time.  The Fourier conjugate of this variable
plays the role of a kinematic skewness in the double parton distributions.  We
discuss the skewness dependence of the relevant hadronic correlation functions.
Using several models, we study the impact of this dependence on extracting
moments of double parton distributions from existing lattice data.}

\emailAdd{Daniel.Reitinger@physik.uni-regensburg.de}

\arxivnumber{2511.15546}

\preprint{DESY-25-167}

\begin{document}

\maketitle

%%%%%%%%%%%%%%%%%%%%%%%%%%%%%%%%%%%%%%%%%%%%%%%%%%%%%%%%%%%%%%%%%%%%%%%%%%%%%%%%

\section{Introduction}

The LHC is presently the leading accelerator searching for New Physics at the
energy frontier and will stay so for many years to come. To make best use of its
data, significant theory efforts are required, in particular in the domain of
strong interaction physics. Among the open challenges in this area is a better
understanding of multiparton interactions, which for particular final states
could mimic or hide signals for New Physics. In multiparton interactions,
several hard processes occur in the same proton-proton collision. The simplest
and most prominent case is double parton scattering, in which two partons from
each proton take part in a hard scattering. A comprehensive overview of this
subject can be found in~\cite{Bartalini:2018qje}.

The initial state of double parton scattering is described by double parton
distributions (DPDs). Whereas single parton distributions (PDFs) have entered
the realm of precision physics, our knowledge of DPDs remains quite fragmentary.
This has several reasons, ranging from their inherent complexity (they depend on
two parton labels, three variables, two scales) to the difficulties of
extracting double parton scattering from experimental measurements. In this
situation, theoretical input that allows us to constrain these functions is of
great value. Beyond their practical relevance for collider physics, DPDs are of
interest in themselves, because they describe quantum correlations inside the
proton that are ``invisible'' in single-parton distributions.

It is natural to ask how lattice QCD can inform us about DPDs, given the long
and successful history of studying single-parton distributions on the lattice
(see \cite{Lin:2017snn,Constantinou:2020hdm} for recent reviews).  In this
context, one may distinguish two different approaches.
\begin{enumerate}
\item Mellin moments of PDFs are related to hadronic matrix elements of local
currents, which are well suited for evaluation on a Euclidean lattice.  This
method has reached a high degree of sophistication regarding the reliability of
results (see Section II in \cite{Constantinou:2020hdm}), but due to
power-divergent operator mixing most studies so far are limited to the lowest
three moments.  While three moments are not sufficient to reconstruct the
dependence of PDFs on their momentum fraction $x$, they may for instance be used
as additional input in PDF fits to experimental data.  Moreover, some of the
moments are related to sum rules and of interest in themselves, such as the
lowest moments of helicity dependent PDFs.

As pointed out in \cite{Diehl:2011yj}, this method can be extended to DPDs by
relating their Mellin moments to the hadronic matrix elements of two currents at
different spatial positions. Computationally this is much more demanding since
it requires the evaluation of four-point functions on the lattice (instead of
three-point functions for PDFs). Nevertheless, we have shown in a series of
studies \cite{Bali:2020mij,Bali:2021gel,Reitinger:2024ulw} that good statistical
signals can be obtained for a variety of correlation functions related to the
lowest Mellin moments of DPDs in the pion or the nucleon.

There is, however, a further complication: to reconstruct the Mellin moments of
DPDs with this method, one needs to evaluate the integral over a variable
$\omega$ that may be regarded as a Ioffe time and physically ranges from
$-\infty$ to $+\infty$.  In lattice simulations one can only access a finite
range in this variable.  In our previous work, we tackled this problem by
introducing a kinematical ``skewness'' variable $\zeta$ that is Fourier
conjugate to $\omega$, such that the Mellin moments of DPDs correspond to $\zeta
= 0$. We then assumed a simple functional form for this $\zeta$ dependence and
determined its parameters by fits to the lattice data.

The purpose of the present work is to take a closer look at this procedure,
exploring a broader range of physically motivated ans\"atze for the $\zeta$
dependence and investigating the corresponding fit results.
\item A variety of methods have been proposed for accessing the $x$ dependence
of PDFs in a more direct manner, see e.g.~section~2 in \cite{Lin:2017snn}.  In
particular, the concepts of quasi-distributions \cite{Ji:2013dva,Ji:2014gla} and
of pseudo-distributions \cite{Radyushkin:2016hsy,Radyushkin:2017cyf} have been
employed in recent years to probe the $x$ dependence of single-parton
quantities, not only for PDFs but also for TMDs (transverse momentum
distributions) and GPDs (generalized parton distributions).

Incidentally, the problem described above --- having to integrate over Ioffe
time for obtaining the quantities of physical interest --- also occurs for
quasi- and pseudo-distributions, where, roughly speaking, a Ioffe time variable
is Fourier conjugate to the momentum fraction $x$.  Also in these cases, lattice
simulations are restricted to a finite range in this variable.  This issue is
under active investigation, see for instance \cite{Chen:2025cxr},
\cite{Xiong:2025obq}, \cite{Dutrieux:2025jed}, and references therein.

An extension of the quasi-distribution method to DPDs has been formulated in
\cite{Jaarsma:2023woo} and \cite{Zhang:2023wea}.  The impressive progress
achieved with this method for single-parton distributions raises hope that it
can be turned into a powerful tool to learn more about DPDs as well.  We shall,
however, not pursue this avenue in the present work.
\end{enumerate}

This work is organized as follows.  In Section~\ref{sec:theory} we recall the
most important properties of DPDs, their extension to finite skewness $\zeta$,
as well as their relation with hadronic matrix elements of two local currents.
In Section~\ref{sec:moment-properties} we derive constraints on the $\zeta$
dependence of Mellin moments of DPDs.  The most important aspects of our lattice
analysis are recalled in Section~\ref{sec:lattice-calc}.  We introduce different
parameterizations for the $\zeta$-dependent moments in Section~\ref{sec:models}
and use these to fit the lattice data in Section~\ref{sec:fit-results}.  We
summarize our findings in Section \ref{sec:summary}.

%%%%%%%%%%%%%%%%%%%%%%%%%%%%%%%%%%%%%%%%%%%%%%%%%%%%%%%%%%%%%%%%%%%%%%%%%%%%%%%%

\section{Theory background}
\label{sec:theory}

In this section, we review basic definitions and properties of DPDs, including
their extension to nonzero skewness $\zeta$ introduced in our previous
lattice studies \cite{Bali:2020mij,Bali:2021gel}.

\subsection{DPDs, skewed DPDs, and their Mellin moments}

DPDs are defined by matrix elements of the product of two twist-two operators.
Throughout this work, we consider only twist-two operators in the color singlet
representation, i.e.\ without open color indices.  For simplicity, we also
restrict ourselves to the case of unpolarized partons, where our lattice data is
most precise.  An extension of our theoretical arguments to polarized partons is
straightforward.

We use light cone coordinates $v^{ \pm}=\left(v^0 \pm v^3\right) / \sqrt{2}$ and
$ \boldsymbol{v}=\left(v^1, v^2\right)$ for a given four-vector $v^\mu$.  The
twist-two operators for an unpolarized quark or antiquark read
\begin{align}
   \label{qqbar-ops}
   \mathcal{O}_q(y, z)
   &=
   \frac{1}{2}\left.
   \bar{q}\! \left(y - \frac{1}{2} z\right)
   \gamma^{+} \,
   q\! \left(y + \frac{1}{2} z\right)
   \right|_{z^{+}=y^{+}=0,\, \tvec{z} = \tvec{0}}
   \,,
   \notag \\[0.3em]
   \mathcal{O}_{\bar{q}}(y, z)
   &=
   - \frac{1}{2}\left.
   \bar{q}\! \left(y + \frac{1}{2} z\right)
   \gamma^{+} \,
   q\! \left(y - \frac{1}{2} z\right)
   \right|_{z^{+}=y^{+}=0,\, \tvec{z} = \tvec{0}}
   \,,
\end{align}
respectively, where we have suppressed the usual Wilson line along the straight
path from $y - \half z$ to $y + \half z$.
The operator $\mathcal{O}_{g}$ for gluons can for instance be found in
Section~2.2 of \cite{Diehl:2011yj}.  It is understood that ultraviolet
divergences are renormalized in the conventional way, using the
$\overline{\text{MS}}$ prescription.  The corresponding scale dependence of
operators and their matrix elements is suppressed for brevity in all of this
work.
The DPD for unpolarized partons $a_1$ and $a_2$ is then defined by
\begin{align}
   \label{ordinary-DPD}
   F_{a_1 a_2}\left(x_1, x_2, y^2\right)
   =
   2 p^{+} \int \mathrm{d} y^{-}
   &
   \int \frac{\mathrm{d} z_1^{-}}{2 \pi}
   \frac{\mathrm{~d} z_2^{-}}{2 \pi} \,
   e^{i \left(x_1^{} z_1^{-} + x_2^{} z_2^{-}\right) p^{+}}
   \nonumber \\[0.2em]
   &
   \times \sideset{}{^{\prime}}\sum_{\lambda}
   \bra{p,\lambda}
   \mathcal{O}_{a_1}(y, z_1) \,
   \mathcal{O}_{a_2}(0, z_2)
   \ket{p,\lambda}
   \,.
\end{align}
Here we have taken the average over the proton polarization $\lambda$, which we
write as $\sum_{\lambda}^{\prime}=\frac{1}{2} \sum_\lambda$. We work in a frame
where the transverse momentum $\tvec{p}$ of the proton is zero. Due to
rotational invariance, the DPD then depends on the transverse distance
$\tvec{y}$ between the two partons only via the Lorentz invariant
\begin{align}
   \label{y2-in-dpd}
   y^2 = - \tvec{y}^2
   \,.
\end{align}
The matrix element in \eqref{ordinary-DPD} has support for both positive and
negative plus-momentum fractions $x_i$ ($i=1,2$), where $x_i > 0$ corresponds to
extracting a parton $a_i$ and $x_i < 0$ corresponds to extracting the charge
conjugate parton $\bar{a}_i$.\footnote{%
For brevity, we use the term ``momentum fraction'' instead of ``plus-momentum
fraction'' from now on.}

The extension of this definition to nonzero skewness $\zeta$ reads
\begin{align}
   \label{skewed-DPD}
   F_{a_1 a_2}\left(x_1, x_2, \zeta, y^2\right)
   =
   2 p^{+} \int \mathrm{d} y^{-}
   &
   e^{-i \zeta y^{-} p^{+}} \int
   \frac{\mathrm{d} z_1^{-}}{2 \pi}
   \frac{\mathrm{~d} z_2^{-}}{2 \pi} \,
   e^{i \left(x_1^{} z_1^{-} + x_2^{} z_2^{-}\right) p^{+}}
   \nonumber \\[0.2em]
   &
   \times \sideset{}{^{\prime}}\sum_{\lambda}
   \bra{p,\lambda}
   \mathcal{O}_{a_1}(y, z_1) \,
   \mathcal{O}_{a_2}(0, z_2)
   \ket{p,\lambda}
   \,.
\end{align}
Compared with \eqref{ordinary-DPD}, we now have an additional phase factor that
involves the Lorenz invariant
\begin{align}
   \label{Ioffe-time}
   \omega = p y = p^+ y^-
   \,,
\end{align}
which we refer to as a Ioffe time.  This variable is Fourier conjugate to the
skewness $\zeta$.  Due to parity and time reversal invariance, one has the
symmetry property
\begin{align}
   \label{DPD-even}
   F_{a_1 a_2}\left(x_1, x_2, \zeta, y^2\right)
   &=
   F_{a_1 a_2}\left(x_1, x_2, - \zeta, y^2\right)
   \,.
\end{align}

\begin{figure}
\begin{center}
\subfloat[\label{support-dpd} $x_i \pm \zeta / 2 > 0$]{
\includegraphics[width=0.49\textwidth]{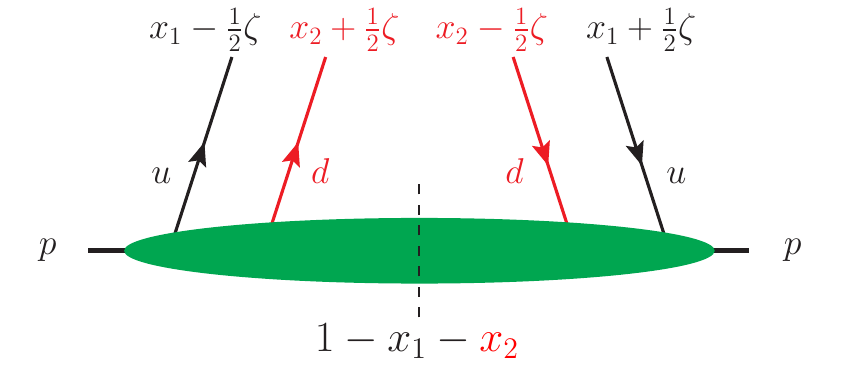}
}
\subfloat[\label{support-central} $\zeta / 2 \pm x_i > 0$]{
\includegraphics[width=0.49\textwidth]{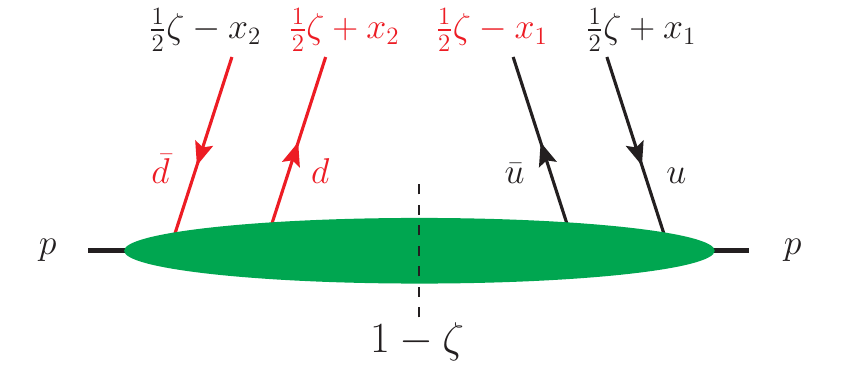}
}
\caption{Illustration of the skewed DPD $F_{u d}\left(x_1, x_2, \zeta,
y^2\right)$ for the cases where (a) all fractions $x_i \pm \zeta / 2$ or (b) all
fractions $\zeta / 2 \pm x_i$ are positive.  Below the vertical dashed line we
give the total momentum fraction carried by the spectator partons.}
\label{zeta-distrib}
\end{center}
\end{figure}

The skewness parameter introduces a difference between the momentum fractions of
the partons in the bra or ket states of the matrix element, viz.\ in the wave
function or the conjugate wave function of the proton. Positive momentum
fractions $x_1 - \zeta/2$ ($x_2 + \zeta/2$) correspond to a parton $a_1$ ($a_2$)
in the wave function, and positive momentum fractions $x_1 + \zeta/2$ ($x_2 -
\zeta/2$) to a parton $a_1$ ($a_2$) in the conjugate wave function, as shown in
Figure~\ref{support-dpd}.  When $x_1 - \zeta/2$ turns from positive to negative
values, $a_1$ in the wave function turns into $\bar{a}_1$ in the conjugate wave
function, with corresponding statements for the other three momentum fractions.
Figure \ref{support-central} shows the configuration with $a_2 \ms \bar{a}_2$ in
the wave function and $a_1 \bar{a}_1$ in the conjugate wave function.

\begin{figure}
\begin{center}
\includegraphics[width=0.9\textwidth]{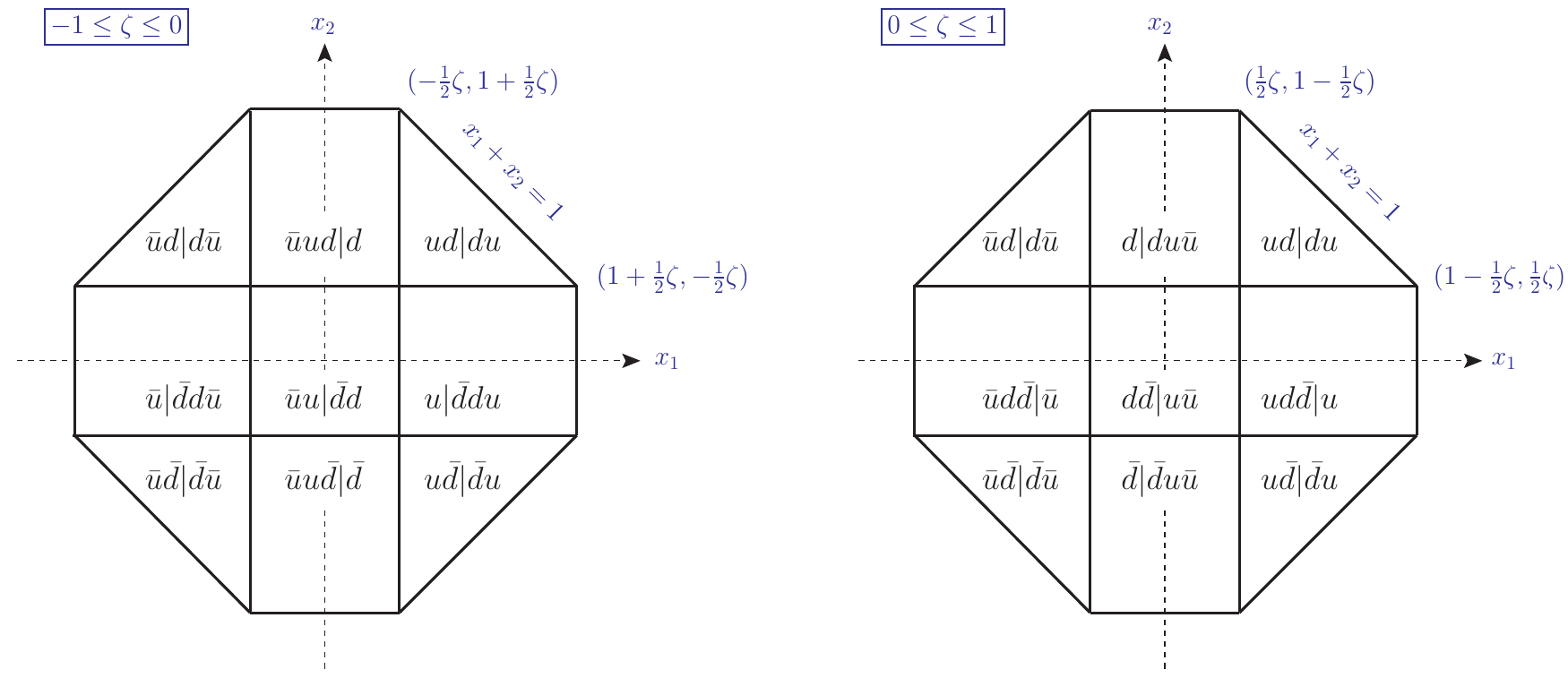}
\caption{Support regions of $F_{u d}\left(x_1, x_2, \zeta, y^2\right)$
in the $(x_1, x_2)$-plane for negative (left) and positive (right) values of
$\zeta$.  For each region we indicate the (anti-)quark content of the wave
function and its complex conjugate. The notation $u | \bar{d} d u$ means that we
have a $u$-quark in the proton wave function and $\bar{d} d u$ in its complex
conjugate.}
\label{dpd-support}
\end{center}
\end{figure}

Using the methods of \cite{Jaffe:1983hp}, one can show that the support property
of the matrix element in \eqref{skewed-DPD} is \cite{Bali:2020mij}
\begin{equation}
\left|x_i \pm \frac{1}{2} \zeta\right| \leq 1
\,,
\quad
\left|x_1\right|+\left|x_2\right| \leq 1
\,,
\quad
|\zeta| \leq 1
\,.
\end{equation}
The different parts of the support region and their partonic interpretation
are visualized in Figure~\ref{dpd-support}.

When taking Mellin moments in the momentum fractions $x_1$ and $x_2$, the non-local twist-two operators in \eqref{ordinary-DPD} and \eqref{skewed-DPD} turn into local currents.  For the lowest Mellin moment we have
\begin{align}
  \label{eq:skewed-inv-mellin-mom-def}
   I_{a_1 a_2}(\zeta, y^2)
   &=
   \int_{-1}^{1} \dd x_1^{} \int_{-1}^{1} \dd x_2^{} \;
   F_{a_1 a_2}(x_1,x_2,\zeta, y^2)
   \notag \\[0.2em]
   &=
   \frac{2}{(p^{+})^2} \int \mathrm{d} \omega \, e^{-i \zeta \omega} \,
   \sideset{}{^{\prime}}\sum_{\lambda}
   \bra{p,\lambda}
   \mathcal{O}_{a_1}\left(y, 0\right)
   \mathcal{O}_{a_2}\left(0, 0\right)
   \ket{p,\lambda}
   \,,
\end{align}
where it is understood that the integrals over $x_1$ and $x_2$ are performed
over the full support region shown in Figure~\ref{dpd-support}.  The symmetry
property \eqref{DPD-even} for skewed DPDs implies
\begin{align}
   \label{moment-even}
   I_{a_1 a_2}(\zeta, y^2)
   &=
   I_{a_1 a_2}(-\zeta, y^2)
   \,.
\end{align}

%%%%%%%%%%%%%%%%%%%%%%%%%%%%%%%%%%%%%%%%

\subsection{The number sum rule}
\label{sec:sumrule}

DPDs fulfill sum rules, which have been proposed in \cite{Gaunt:2009re} and
proven in \cite{Gaunt:2012tfk,Diehl:2018kgr}. In the context of our lattice
calculations, only the number sum rule is relevant.  For the two-quark DPD $F_{q
q'}$ it can be written as
\begin{align}
   \label{eq:dpd-sr}
   &
   2\pi \int_{y_{\text{cut}}}^\infty \dd y\, y \,
   \int_{-1+|x_1|}^{1-|x_1|} \dd x_2 \;
   F_{q q^\prime}(x_1,x_2, y^2)
   \nonumber \\[0.2em]
   & \qquad
   =
   \begin{cases}
      \left( N_{q^\prime} - \delta_{q q^\prime} \right) f_q(x_1)
      + \mathcal{O}\bigl( \alpha_s \bigr)
      + \mathcal{O}(\Lambda^2 y_{\text{cut}}^2)
      & \text{ for } x_1 > 0 \,,
      \\[0.3em]
      - \left( N_{q^\prime} + \delta_{q q^\prime} \right) f_{\bar{q}}(- x_1)
      + \mathcal{O}\bigl( \alpha_s \bigr)
      + \mathcal{O}(\Lambda^2 y_{\text{cut}}^2)
      & \text{ for } x_1 < 0 \,,
   \end{cases}
\end{align}
where $f_q$ and $f_{\bar{q}}$ are the usual quark and antiquark distributions
and $N_{q^\prime}$ is the number of valence quarks $q^\prime$.  To obtain this
form, we have used the relation $F_{\bar{q} q'}(x_1, x_2, y^2) = - F_{q
q'}(-x_1, x_2, y^2)$ and its counterpart for the second parton, which follow
from the relation between quark and antiquark operators in \eqref{qqbar-ops}.

As discussed in \cite{Diehl:2018kgr}, the lower cutoff on the integral over $y$
is necessary to deal with the singular $1/y^2$ behavior of $F_{q q^\prime}$ at
$y \rightarrow 0$.  This behavior is due to perturbative splitting and comes
with at least one factor of $\alpha_s$.
The last term in \eqref{eq:dpd-sr} denotes power corrections, with $\Lambda$
representing a typical hadronic scale.  One should therefore take
$y_{\text{cut}}$ sufficiently small.  We use $y_{\mathrm{cut}} = b_0/\mu$, where
$\mu$ is the renormalization scale of the DPD and $b_0 = 2e^{-\gamma} \approx
1.12$ with the Euler-Mascheroni constant $\gamma$.  Taking $\mu = 2 \gev$ we
get
\begin{align}
   \label{y-cut}
   y_{\text{cut}} & \approx 0.11 \fm
   .
\end{align}
In Section~\ref{sec:y-dependence} we comment on the dependence of the integral
over $y$ on its boundaries.

When considering Mellin moments in both momentum fractions, we want to integrate
\eqref{eq:dpd-sr} over $x_1$.  This is only well defined for $q\neq q^\prime$,
because the corresponding expression for $q = q^\prime$ includes sea quark
contributions that lead to divergent results when integrated over all momentum
fractions. Taking the parton combination $u d$ in \eqref{eq:dpd-sr} and using
the number sum rule for ordinary PDFs, we get
\begin{align}
   \label{eq:sr-proton}
   S_{u d}(y_{\text{cut}})
   &=
   2\pi \int_{y_{\text{cut}}}^\infty \dd y\, y \,
   I_{u d}(\zeta=0,y^2)
   =
   2\pi \int_{y_{\text{cut}}}^\infty \dd y\, y \,
   \int_{-1}^1 \dd x_1 \int_{-1}^1 \dd x_2 \,
   F_{u d}(x_1,x_2, y^2)
   \notag \\[0.2em]
   &=
   2 + \Op(\alpha_s^2) + \Op(\Lambda^2 y_{\text{cut}}^2)
   \,.
\end{align}
Notice that the power of $\alpha_s$ on the r.h.s.\ is higher than in the generic
relation \eqref{eq:dpd-sr}, because perturbative splitting for the parton
combination $u d$ starts only at order $\alpha_s^2$.

%%%%%%%%%%%%%%%%%%%%%%%%%%%%%%%%%%%%%%%%

\subsection{Two-current matrix elements}

The relation between two-current matrix elements and DPDs has been worked out in
detail in \cite{Bali:2020mij,Bali:2021gel}. For the present work, we need the
correlation function
\begin{align}
  \label{eq:mat-els}
   M^{\mu_1 \mu_2}_{q_1 q_2, V V}(p, y)
   &=
   \sideset{}{^{\prime}}\sum_{\lambda}
   \bra{p,\lambda}
      J^{\mu_1}_{q_1, V}(y)\, J^{\mu_2}_{q_2, V}(0)
   \ket{p,\lambda}
\end{align}
of two quark vector currents
\begin{align}
   \label{eq:local-ops}
   J_{q, V}^\mu(y) &= \bar{q}(y) \ms \gamma^\mu\ms q(y)
   \,.
\end{align}
This matrix element can be decomposed in terms of Lorentz invariant functions,
which depend on the Lorentz scalars $\omega = p y$ and $y^2$ only.  Symmetrizing
the tensor \eqref{eq:mat-els} and subtracting its trace, we get
\begin{align}
   \label{eq:tensor-decomp}
   &
   \tfrac{1}{2} \bigl(
      M^{\mu\nu}_{q_1 q_2, V V} + M^{\nu\mu}_{q_1 q_2, V V} \bigr)
   - \tfrac{1}{4} g^{\mu\nu} g_{\alpha\beta}
      M^{\alpha\beta}_{q_1 q_2, V V}
   \notag \\[0.2em]
   & \qquad =
   u_{V V,A}^{\mu\nu}\, A_{q_1 q_2}^{}(\omega,y^2)
   + u_{V V,B}^{\mu\nu}\, m^2\ms B_{q_1 q_2}^{}(\omega,y^2)
   + u_{V V,C}^{\mu\nu}\, m^4\ms C_{q_1 q_2}^{}(\omega,y^2)
\end{align}
with tensors $u^{\mu\nu}$ given in Section~2.3 of \cite{Bali:2021gel}.  Comparing
this with the matrix element representation \eqref{eq:skewed-inv-mellin-mom-def}
of the Mellin moment $I_{q_1 q_2}$, one finds
\begin{align}
   \label{mellin-inv-fct}
   I_{q_1 q_2}(\zeta, y^2)
   &=
   \int_{-\infty}^{\infty} \dd\omega\, e^{-i\zeta \omega}\,
      A_{q_1 q_2}(\omega,y^2) \,,
   \\[0.3em]
   \label{inv-fct-mellin}
   A_{q_1 q_2}(\omega, y^2)
   &=
   \frac{1}{2\pi}\int_{-1}^{1} \dd\zeta\, e^{i\zeta \omega}\,
      I_{q_1 q_2}(\zeta,y^2)
   \,,
\end{align}
where in the second line we have used the limited support of $I_{q_1 q_2}$ in
$\zeta$.  These relations imply that $A_{q_1 q_2}(\omega, y^2)$ is even in
$\omega$, because $I_{q_1 q_2}(\zeta, y^2)$ is even in $\zeta$.

The distance $y$ between the twist-two operators in DPDs is always spacelike, as
can be seen in \eqref{y2-in-dpd}.  One also gets spacelike $y$ if one sets $y^0
= 0$ in the two-current correlators \eqref{eq:mat-els}, which can then be
evaluated in Euclidean space-time.  From this one can extract the invariant
functions $A_{q_1 q_2}(\omega, y^2)$, which appear in  \eqref{mellin-inv-fct}.
We thus have the situation announced in the introduction.  Our primary physics
interest is in the Mellin moments $I_{q_1 q_2}(\zeta = 0, y^2)$ of DPDs at zero
skewness, since these DPDs appear in double parton scattering.  To evaluate the
Mellin moments from correlation functions accessible to lattice simulations, we
need to perform an integral over the Ioffe time $\omega$.

%%%%%%%%%%%%%%%%%%%%%%%%%%%%%%%%%%%%%%%%%%%%%%%%%%%%%%%%%%%%%%%%%%%%%%%%%%%%%%%%

\section{Properties of Mellin moments}
\label{sec:moment-properties}

Little is known about the functional form of the Mellin moments $I_{a_1
a_2}(\zeta, y^2)$, apart from the symmetry constraint that they are even
functions of~$\zeta$.  In our previous analysis \cite{Bali:2021gel} of lattice
results, we assumed a polynomial form in $\zeta$. Whilst this ansatz produced
stable fits to the data for $A_{a_1 a_2}(\omega, y^2)$, we now show that it
misses an important boundary condition and at the same is too restrictive.
Specifically, we argue that
\begin{enumerate}
\item $I_{a_1 a_2}(\zeta, y^2)$ should vanish rather fast in the limit $\zeta
\to 1$,
\item in general, $I_{a_1 a_2}(\zeta, y^2)$ is not an analytic function of
$\zeta$ at the points $\zeta = 0$ and $\zeta = 1$.
\end{enumerate}

Throughout this section will assume $\zeta \ge 0$ for ease of notation,
bearing in mind that $I_{a_1 a_2}(\zeta, y^2)$ is even in $\zeta$.

The first property follows from a rather general physics consideration.  As
$\zeta \to 1$, the support of skewed DPDs in $x_1$ and $x_2$ reduces to the
central square $|x_1|, |x_2| \le \half \zeta$ in Figure \ref{dpd-support}.  The
corresponding parton configuration is the one shown in Figure
\ref{support-central}, where the two extracted partons together carry a
longitudinal fraction $\zeta$ of the proton momentum, leaving a fraction $1 -
\zeta$ for the remaining spectator partons.  More specifically, for the parton
combinations $(a_1 a_2) = (u d), (u u), (d d)$ accessible in our lattice data,
the moment $I_{a_1 a_2}$ at $\zeta = 1$ describes the extraction of a
color-singlet $u \bar{u}$ or $d \bar{d}$ pair that carries the full longitudinal
proton momentum, with zero longitudinal momentum of the remaining partons (which
include two or three ``valence quarks'' if one regards the proton as a system
made from three valence quarks plus sea quarks and gluons). We find it plausible
that the probability for such configurations is zero and that the approach to
zero should be fast as $\zeta \to 1$, to a similar extent as ordinary $u$ or $d$
quark PDFs quickly approach zero as their momentum fraction $x$ tends to $1$.

Somewhat unfortunately, the region $\zeta \approx 1$ constrained by this
argument is far away from the limit $\zeta \to 0$ that brings us to ordinary
(non-skewed) DPDs and the physics of double parton scattering.  In the following
subsection we therefore present an argument for the second property stated
above.

%%%%%%%%%%%%%%%%%%%%%%%%%%%%%%%%%%%%%%%%

\subsection{Lessons from the small-distance limit}
\label{sec:small-y-limit}

\begin{figure}[bh]
\begin{center}
\subfloat[\label{split-lo}]{
\includegraphics[width=0.32\textwidth]{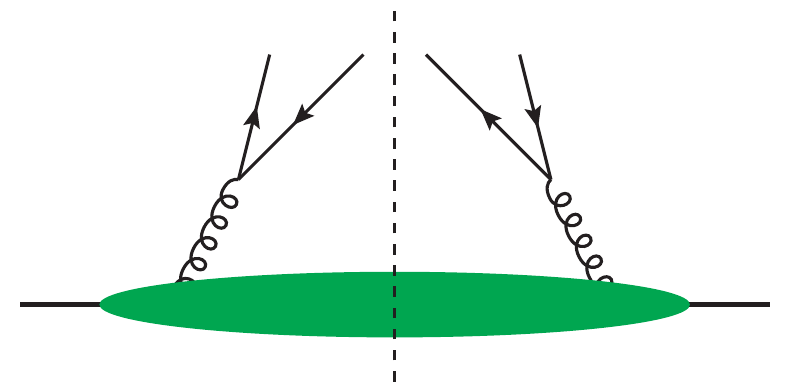}
}
\subfloat[\label{split-nlo}]{
\includegraphics[width=0.32\textwidth]{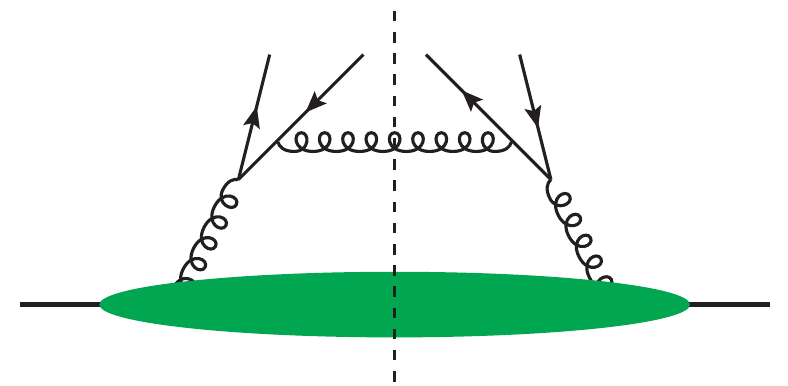}
}
\subfloat[\label{split-int}]{
\includegraphics[width=0.32\textwidth]{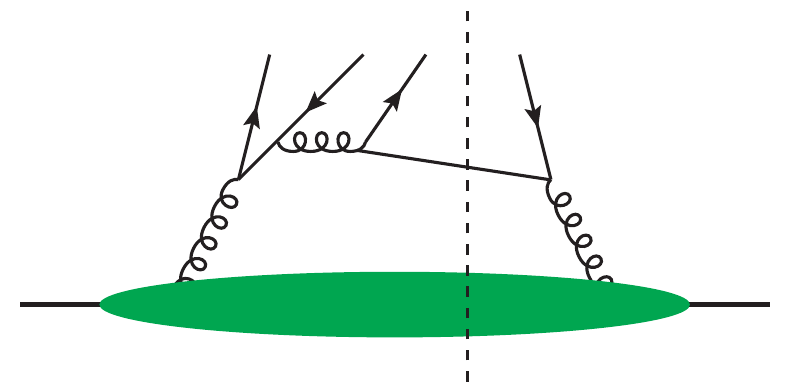}
}
\caption{Graphs for the perturbative splitting mechanism of DPDs at leading (a)
or next-to-leading (b, c) order in $\alpha_s$.  Graph (c) only contributes for
$\zeta \neq 0$.}
\label{fig:splitting}
\end{center}
\end{figure}

As shown in \cite{Diehl:2011yj,Diehl:2017kgu}, ordinary DPDs in the limit of
small interparton distance $y$ are dominated by the mechanism in which the two
extracted partons originate from the splitting of a single parton.  One can then
compute the DPD in terms of ordinary PDFs and perturbative splitting kernels.
The arguments for this type of factorization extend to the case of nonzero
skewness $\zeta$.  Corresponding Feynman graphs are shown in Figure
\ref{fig:splitting}.

At leading order (LO) in $\alpha_s$, the graph in Figure \ref{split-lo} can
contribute to DPDs in the regions given by the triangles $|x_1| \ge \half
\zeta$, $|x_2| \ge \half \zeta$ in the $(x_1, x_2)$ plane.  Since the
quark-antiquark pairs on either side of the final-state cut are coupled to a
color octet in this graph, it does not populate the central square $|x_1| \le
\half \zeta$, $|x_2| \le \half \zeta$, which corresponds to extracted
quark-antiquark pairs in a color singlet.  By contrast, the next-to-leading
order (NLO) graph in Figure  \ref{split-nlo} contributes to both the triangles
and the central square in the support octagon.  NLO graphs such as the one in
\ref{split-int} populate the outer rectangles where $|x_1| \ge \half \zeta$ and
$|x_2| \le \half \zeta$ or vice versa.

We will shortly discuss the contribution of the LO graph for $g \to q \bar{q}$
splitting to the Mellin moment $I_{q q}$.  Before doing so, we should point out
that this contribution has no direct connection to the lattice data analyzed in
this work.  For one thing, distances $y$ small enough for a perturbative
treatment correspond to less than one or two units of the lattice spacing $a
\approx 0.0856 \fm$, and at such distances we must expect significant
discretization effects. Moreover, the perturbative graphs in
Figure~\ref{fig:splitting} correspond to the lattice contraction $S_2$ in Figure
\ref{fig:graphs}, which will be excluded from our numerical analysis for the
reasons given in Section~\ref{sec:lattice} .  However, for showing that it is
\emph{not a general property} of the Mellin moments $I_{a_1 a_2}(\zeta, y^2)$ to
be analytic functions of $\zeta$ at $\zeta = 0$ or $\zeta = 1$ it is sufficient
to consider a restricted region of $y$, a restricted set of contributing graphs
or lattice contractions, and a particular combination of the partons $a_1$ and
$a_2$.

As shown in appendix \ref{app:splitting}, the contribution of Figure
\ref{split-lo} to the lowest Mellin moment of $F_{q q}$ reads
\begin{align}
   \label{splitting-lowest-moment}
   I_{q q}(\zeta, y^2)
   &=
   - \frac{2}{\pi |y^2|} \, \frac{\alpha_s}{2 \pi} \, T_F \,
   \int_{\zeta}^1 \dd z \, f_g(z) \;
   \frac{1}{3} \biggl( 1 - \frac{\zeta}{z} \biggr)^2 \,
   \biggl( 2 +  \frac{\zeta}{z} \biggr) \,.
\end{align}
where $T_F = 1/2$ and $f_g(z)$ is the gluon distribution in the proton.
Assuming a behavior
\begin{align}
   \label{gluon-PDF}
   f_g(z) & \sim
   z^{- 1 + \alpha} \, (1 - z)^\beta
   &
   \text{with } -1 < \alpha < 0
\end{align}
and $\beta$ much larger than unity, one finds that
\begin{align}
   \label{splitting-large-zeta}
   I_{q q}(\zeta, y^2)
   & \sim
   \int_{\zeta}^1 \dd z \, (1 - z)^\beta \, (z - \zeta)^2
   \sim
   (1 - \zeta)^{3 + \beta}
   &
   \text{for } \zeta \to 1
   \,.
\end{align}
This provides a concrete example for our expectation that the Mellin moment
vanishes rather fast as $\zeta \to 1$.

With the condition on $\alpha$ in \eqref{gluon-PDF}, the integral in
\eqref{splitting-lowest-moment} diverges as $\zeta \to 0$, and one finds a
singular behavior
\begin{align}
   \label{splitting-small-zeta}
   I_{q q}(\zeta, y^2)
   & \sim
   \zeta^\alpha
   &
   \text{for } \zeta \to 0
   \,.
\end{align}
This is closely related to fact the lowest Mellin moment $\int_0^1 \dd z\, f_g(z)$
of the gluon PDF is divergent as well.  It is also in line with the observation
we made in Section~\ref{sec:sumrule}, namely that the DPD number sum rule
for equal quark flavors gives infinity when integrated over \emph{both} momentum
fractions.

It is instructive to consider in addition the Mellin moment with weight factor
$x_1 x_2$, for which the leading order perturbative splitting gives
\begin{align}
   \label{splitting-higher-moment}
   I^{(2, 2)}_{q q}(\zeta, y^2)
   &=
   \int_{-1}^{1} \dd x_1^{} \, x_1 \int_{-1}^{1} \dd x_2^{} \, x_2 \;
   F_{q q}(x_1,x_2,\zeta, y^2)
   \notag \\
   &=
   \frac{2}{\pi |y^2|} \, \frac{\alpha_s}{2 \pi} \, T_F \,
   \int_{\zeta}^1 \dd z \, z^2 \ms f_g(z) \;
   \frac{1}{60} \biggl( 1 - \frac{\zeta}{z} \biggr)^2 \,
   \biggl( 6 +  \frac{12 \zeta}{z} - \frac{2 \zeta^2}{z^2}
      -  \frac{\zeta^3}{z^3} \biggr) \,.
\end{align}
One easily finds that for a gluon distribution of the form \eqref{gluon-PDF}
this has same behavior at $\zeta \to 1$ as $I_{q q}(\zeta, y^2)$ in
\eqref{splitting-large-zeta}.  Thanks to the factor $z^2$ in the integrand, it
has a finite value at $\zeta = 0$, but it is non-analytic at that point.

In summary, the above example shows that a non-analytic behavior of PDFs at
momentum fraction $z=0$ and $z=1$ induces a non-analytic behavior of Mellin
moments of DPDs at skewness $\zeta=0$ and $\zeta=1$ if the distance $y$ is
sufficiently small.  It is difficult to imagine how the same Mellin moments
could be analytic in $\zeta$ at larger values of $y$, where the perturbative
splitting formula for DPDs does not hold.

%%%%%%%%%%%%%%%%%%%%%%%%%%%%%%%%%%%%%%%%%%%%%%%%%%%%%%%%%%%%%%%%%%%%%%%%%%%%%%%%

\section{Lattice calculation}
\label{sec:lattice-calc}

In this section we recall the basic steps for extracting the invariant
functions $A_{q_1 q_2}(\omega, y^2)$ from lattice simulations.  For details and
plots we refer to the reader to our previous work~\cite{Bali:2021gel}.

%%%%%%%%%%%%%%%%%%%%%%%%%%%%%%%%%%%%%%%%

\subsection{Two-current matrix elements}

\begin{figure}
\begin{center}
\includegraphics[scale=1]{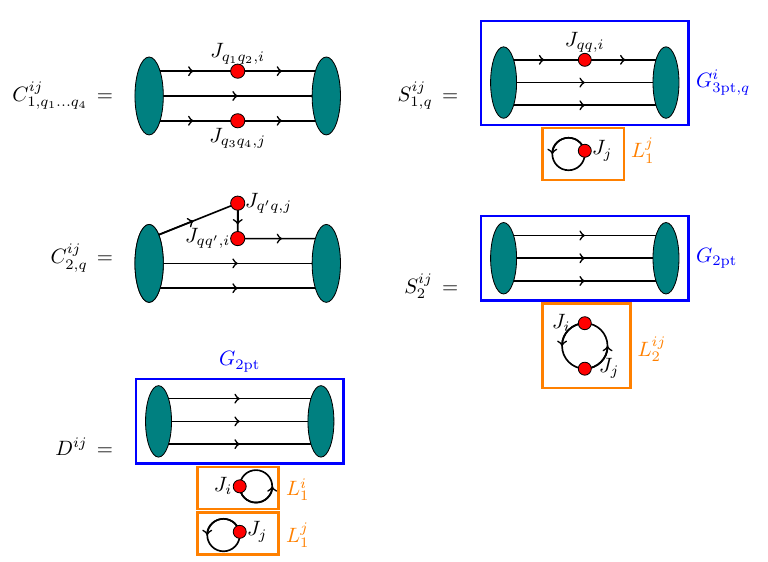}
\caption{\label{fig:graphs} Illustration of the five kinds of Wick contractions
(graphs) contributing to a four-point function of a baryon. The explicit
contributions for the graphs $C_1$, $C_2$ and $S_1$ depend on the quark flavor
of the current insertions (red points). In the case where all quark flavors have
the same mass, $C_2$ only depends on the flavors of the two propagators
connected to the source or the sink. These flavors have to be the same for
proton-proton matrix elements. For the graphs $S_1$, $S_2$ and $D$ we also
indicate the parts connected to the proton source and sink, i.e.
$G_{3\mathrm{pt}}$ and $G_{2\mathrm{pt}}$ (blue), as well as the disconnected
loops $L_1$ and $L_2$ (orange).  The indices $i$ and $j$ specify the currents
and are both equal to $V$ in our context.  [Figure taken from
Ref.~\protect\cite{Bali:2021gel}]}
\end{center}
\end{figure}

In order to evaluate the two-current matrix element \eqref{eq:mat-els} on the
lattice, we switch from Minkowski space to Euclidean space.  As is customary,
the corresponding time component is labeled by the index 4 instead of 0.
We then define the following four-point function for the proton:
\begin{align}
   C^{V V,\mvec{p}}_{\mathrm{4pt},\ms q q'}(\mvec{y},t,\tau)
   &=
   a^6 \, \sum_{\mvec{z}^\prime,\mvec{z}}
   e^{-i\mvec{p}(\mvec{z}^\prime-\mvec{z})}\
   \left\langle \tr \left\{
      P_+ \mathcal{P}(\mvec{z}^\prime,t)\ J_{q,V}(\mvec{y},\tau)\
      J_{q',V}(\mvec{0},\tau)\ \overline{\mathcal{P}}(\mvec{z},0)
   \right\} \right\rangle
   \,,
   \label{eq:4ptdef}
\end{align}
as well as the two-point function
\begin{align}
   C^{\mvec{p}}_{\mathrm{2pt}}(t)
   &=
   a^6 \, \sum_{\mvec{z}^\prime,\mvec{z}}
   e^{-i\mvec{p}(\mvec{z}^\prime-\mvec{z})}\
   \left\langle \tr \left\{
      P_+ \mathcal{P}(\mvec{z}^\prime,t)\
      \overline{\mathcal{P}}(\mvec{z},0) \right\}
   \right\rangle
   \,.
   \label{eq:2ptdef}
\end{align}
The phase factor $e^{-i\mvec{p}(\mvec{z}^\prime-\mvec{z})}$ selects proton
states with the desired three-momentum $\mvec{p}$. The operator
$J_{q, V}(\mvec{y},\tau)$ is the Euclidean version of the vector current
\eqref{eq:local-ops}, and $P_+$ projects onto positive parity. The interpolating
operators $\mathcal{P}(\mvec{z}^\prime,t)$ $\left(
\overline{\mathcal{P}}(\mvec{z},0)  \right)$ annihilate (create) a baryon with
the quantum numbers of the proton, i.e.\ spin $J=1/2$ and isospin $I=1/2$:
\begin{align}
   \overline{\mathcal{P}}(\mvec{x},t)
   &=
   \left.\epsilon_{a b c}\
      \left[
         \bar{u}_a(x)\ C \gamma_5\ \bar{d}_b^{\,T}(x)
      \right] \bar{u}_c(x)
   \right|_{x^4=t} \,,
   \nonumber \\[0.2em]
   \mathcal{P}(\mvec{x},t)
   &=
   \left.\epsilon_{a b c}\ u_a(x)
      \left[
         u_b^T(x)\ C \gamma_5\ d_c(x)
      \right]
   \right|_{x^4=t}
   \,,
\label{eq:interpdef}
\end{align}
where $C$ is the charge conjugation matrix in spinor space. The traces in
\eqref{eq:4ptdef} and \eqref{eq:2ptdef} are taken with respect to the open
spinor indices introduced by the quark fields $u_a$ and $\bar{u}_c$. We denote
the time separation between the source and current insertions as $\tau$. The
ground state matrix element \eqref{eq:mat-els} can be extracted by considering
large time separations:
\begin{align}
   \left. \vphantom{\sum}
   M_{q q', V V}(p,y)
   \right|_{y^0 = 0}
   &=
   2 V \sqrt{m^2 + \mvec{p}^2} \,
   \left.
      \frac{C^{V V,\mvec{p}}_{\mathrm{4pt},\ms q q'}(\mvec{y},t,\tau)}
      {C^{\mvec{p}}_\mathrm{2pt}(t)}
   \right|_{0 \ll \tau \ll t}
   \,.
\label{eq:4pt-2pt-ratio}
\end{align}
The four-point function \eqref{eq:4ptdef} receives contributions from several
Wick contractions of the quark fields, which are shown in Figure
\ref{fig:graphs}. Different combinations of these contractions contribute to the
matrix elements:
\begin{align}
M_{u u, i j}(p,y) \,\big|_{y^0 = 0}
& =
C_{1, u u u u}^{i j, \mvec{p}}(\mvec{y})
+C_{2, u}^{i j, \mvec{p}}(\mvec{y})+C_{2, u}^{j i, \mvec{p}}(-\mvec{y})
\notag \\
& \quad
+S_{1, u}^{i j, \mvec{p}}(\mvec{y})
+S_{1, u}^{j i, \mvec{p}}(-\mvec{y})+S_2^{i j, \mvec{p}}(\mvec{y})
+D^{i j, \mvec{p}}(\mvec{y})
\,,
\notag \\[0.3em]
M_{dd, i j}(p,y) \,\big|_{y^0 = 0}
& =
C_{2, d}^{i j, \mvec{p}}(\mvec{y})+C_{2, d}^{j i, \mvec{p}}(-\mvec{y})
\notag \\
& \quad
+S_{1, d}^{i j, \mvec{p}}(\mvec{y})+S_{1, d}^{j i, \mvec{p}}(-\mvec{y})
+S_2^{i j, \mvec{p}}(\mvec{y})+D^{i j, \mvec{p}}(\mvec{y})
\,,
\notag \\[0.3em]
M_{u d, i j}(p,y) \,\big|_{y^0 = 0}
&=
   C^{i j,\mvec{p}}_{1,u u d d}(\mvec{y}) +
   S^{i j,\mvec{p}}_{1,u}(\mvec{y}) +
   S^{j i,\mvec{p}}_{1,d}(-\mvec{y}) +
   D^{i j,\mvec{p}}(\mvec{y})
\,,
\label{eq:phys_me_decomp}
\end{align}
where $i = j = V$ in the present context.
A detailed derivation and listing of all matrix elements as well as the
method of computation of the Wick contractions are given in \cite{Bali:2021gel}.

The bare lattice operator $J^{\mathrm{latt}}_{q,V}(y)$ is renormalised by the
factor $Z_V$, which includes the conversion to the
$\overline{\mathrm{MS}}$-scheme:
\begin{align}
\label{eq:latt_op_ren}
J_{q,V}^{\overline{\mathrm{MS}}}(y)
=
Z_V^{} \ms J_{q,V}^{\mathrm{latt}}(y)
\,.
\end{align}
Since the vector current has a zero anomalous dimension, $Z_V$ is scale
independent.

%%%%%%%%%%%%%%%%%%%%%%%%%%%%%%%%%%%%%%%%

\subsection{Lattice setup}
\label{sec:lattice}

In this study we reuse the data that have been generated in the context of
\cite{Bali:2021gel}. The simulation includes 990 configurations of the H102
ensemble provided by the CLS Collaboration \cite{Bruno:2014jqa}, with a lattice
volume of $32^3\times96$ and open boundary conditions in time direction, where
the Lüscher-Weisz gauge action and $n_f=2+1$ Sheikoleslami-Wohlert fermions have
been employed. Further details are given in Table \ref{tab:cls}.\footnote{%
We take $a = 0.0856 \fm$ for the lattice size in order to be consistent with our
previous study \protect\cite{Bali:2021gel}.  More recent determinations
\cite{RQCD:2022xux,Conigli:2025qvh} differ from this value by up to 1\%, which
is of no practical relevance in the present context.}

\begin{table}
\begin{center}
\begin{tabular}{cccccccccc}
\hline
\hline
id & $\beta$ & $a [\fm]$  & $L^3 \times T$ & $\kappa_{l}$, $\kappa_{s}$
   & $m_{\pi,K} [\mev]$ & $m_\pi L a$ & configs. \\
\hline
H102 & $3.4$ & $0.0856$ & $32^3 \times 96$ & $0.136865$, $0.136549339$
   & $355$, $441$ & $4.9$ & $2037$ \\
\hline
\hline
\end{tabular}
\caption{\label{tab:cls} Details of the CLS ensemble used for the present study.
\rev{Only 990 configurations were used to generate the four-point functions we
are analyzing.}}
\end{center}
\end{table}

The renormalization factor for the vector current has been determined in
\cite{Bali:2020isn} and can be found in table XXIX therein. For $\beta=3.4$ it
is equal to $Z_V = 0.7115$.

The nucleon source and sink were improved by the application of momentum
smearing \cite{Bali:2016lva} with $250$ smearing iterations. To avoid artifacts
stemming from the open boundary conditions in time, the source was chosen at
$t_{\text{src}} = 48 \ms a$. The sink was placed depending on the momentum as
specified in table \ref{tab:t_src_snk}. The operators were inserted on all time
slices between source and sink for the $C_1$ contraction and at $\tau = t/2$ for
$C_2$. Excited states were found to be sufficiently suppressed. The nucleon mass
was determined to be $m_N=1.1296(75) \gev$.

Compared to the connected contributions $C_1$ and $C_2$, the disconnected
terms $S_1$ and $S_2$ were found to be small in the considered region of
distances $\mvec{y}$ (see below) and, thus, are not considered further here. The
doubly disconnected contribution $D$ exhibits huge statistical uncertainties,
making a determination of this contribution unfeasible at this point in time.
Following the argument in \cite{Bali:2018nde}, we assume that the ratio $D :
S_1$ is similar in size to the ratio $S_1 : C_1$, which allows us to neglect the
$D$ contribution as well.

In the numerical studies of the present work, we limit ourselves to the matrix
element for the flavor combination $u d$. The other two combinations, $u u$ and
$d d$, include the $C_2$ graph, which involves a propagator connecting the two
current insertions and has considerably larger statistical errors than $C_1$.
This would reduce the stability of the fits we will perform. With the lattice
data we have at hand, we therefore refrain from performing fits for the
equal-flavor combinations.

\begin{table}
\begin{center}
\begin{tabular}{c c c}
\hline
\hline
$\mvec{P}$ & $|\mvec{p}|[\gev]$ & $(t_{\text{snk}} - t_{\text{src}}) [a]$
   \rule{0pt}{1.2em} \\
\hline
$(0,0,0)$ & $0$ & 12 \\
\hline
$(-1,-1,-1)$ & $0.78$ & 10 \\
\hline
\rev{$(-2, -2, -2),\ (2, 2, -2),\ (2, -2, 2),\ (-2, 2, 2)$}  & $1.57$ & 10 \\
\hline
\hline
\end{tabular}
\caption{\label{tab:t_src_snk} Summary of momenta and source-sink separations in
our lattice data.}
\end{center}
\end{table}

With the time separation $y^0$ between the two currents being zero, the
Lorentz invariants are given by
\begin{align}
   y^2 &= - \mvec{y}^2
   \,,
   &
   \omega &= - \mvec{p} \mvec{y}
   \,,
\end{align}
so that the Ioffe time is limited to
\begin{align}
   \label{omega-bound}
   | \omega |
   & \le
   | \mvec{p} | \, | \mvec{y} |
   \,.
\end{align}
In order to obtain large $\omega$, one hence needs large proton momenta. For
reasons given below, these are chosen in the direction of the lattice diagonals:
\begin{align}
   \label{lattice-momenta}
   \mvec{p} &= \frac{2\pi\mvec{P}}{L a}
   \,,
   &
   \mvec{P} &= (\pm k,\pm k,\pm k)
   \,,
   &
   k = 0, 1, 2
   \,,
\end{align}
where $L=32$ is the spatial lattice extent and $a=0.0856 \fm$ is the lattice
spacing. This leads to the physical momenta listed in table \ref{tab:t_src_snk}.
As the signal gets worse for larger momenta, the calculation was repeated four
times for $k=2$, using four of the eight possible choices of diagonals.  With
the momenta \eqref{lattice-momenta} the bound \eqref{omega-bound} becomes
$|\omega |\le 2.72 \, |\mvec{y}| \, \big/ (4 a)$.

Let us note that the improved interpolating operators suggested in
\cite{Zhang:2025hyo} should allow us in the future to access significantly
larger momenta and thus larger $\omega$.

%%%%%%%%%%%%%%%%%%%%%%%%%%%%%%%%%%%%%%%%

\subsection{Data analysis}
\label{sec:analysis}

Having extracted the tensor components of the two-current matrix element from
\eqref{eq:4pt-2pt-ratio}, we determine the invariant function $A_{q_1
q_2}(\omega, y^2)$ by solving the system of linear equations in
\eqref{eq:tensor-decomp}. Since this system is overdetermined, we solve it by
$\chi^2$ minimization.  We can test for lattice
artifacts from discretization at small $y$ and from finite-size effects at large
$y$ by investigating to which extent the values of $A_{q_1 q_2}$ fitted to the
continuum relation \eqref{eq:tensor-decomp} depend only on the invariants $y^2$
and $\omega$.  Specifically, we tested for a dependence on the direction of
$\mvec{y}$ (anisotropy) and on the size of the proton momentum (frame
dependence) when $y$ and $\omega$ were kept fixed.\footnote{%
For ease of writing, we henceforth write $y$ for the distance $\sqrt{-y^2} =
|\mvec{y}|$ when there is no risk of confusion.}
We found in \cite{Bali:2021gel} that lattice artifacts were smallest for
$\mvec{y}$ close to the space diagonals of the lattice and hence limit our data
extraction to distance vectors for which the angle $\theta$ between $\mvec{y}$
and the nearest lattice diagonal satisfies
\begin{align}
   \label{theta-cut}
   \cos{\theta} > 0.9 \,.
\end{align}.
To maximize
the reach in $\omega$, we consequently took proton momenta along the lattice
diagonals as specified in \eqref{lattice-momenta}.

To limit discretization and finite size effects, we will consider data only in
the region
\begin{align}
   \label{y-selected-range}
   4 a
   & \le y \le
   16 a
   \,,
\end{align}
which in physical units approximately corresponds to $0.34 \fm < y < 1.37 \fm$.
The upper limit on $y$ corresponds to half the lattice size.  Is not too
relevant in the sense that for $y$ close to $16 a$  statistical uncertainties
become rather large, so that this region has little impact on our fits.

%%%%%%%%%%%%%%%%%%%%%%%%%%%%%%%%%%%%%%%%%%%%%%%%%%%%%%%%%%%%%%%%%%%%%%%%%%%%%%%%

\section{Modeling the lowest Mellin moment}
\label{sec:models}

Our strategy for evaluating Mellin moments $I_{q_1 q_2}(\zeta, y^2)$ is to make
a model (in the sense of an ansatz) for their functional dependence.  For a
given model the Fourier transform \eqref{inv-fct-mellin} can readily be
evaluated.  The parameters of the model are then fitted to the data on $A_{q_1
a_2}(\omega, y^2)$, which have been extracted from our lattice simulations as
described in the preceding section.

To obtain stable fits, we proceed in two steps.  We first make an ansatz for the
function
\begin{align}
   \label{K-def}
   K_{q_1 q_2}(y^2) = A_{q_1 q_2}(\omega = 0, y^2)
\end{align}
and fit its parameters to the data for $\omega = 0$, which has by far the
highest statistical precision because it includes the case of zero proton
momentum.  With this function fixed, we perform fits to the normalized
quantity
\begin{align}
   \label{A-hat-def}
   \hat{A}_{q_1 q_2}(\omega, y^2)
   &=
   \frac{A_{q_1 q_2}(\omega, y^2)}{A_{q_1 q_2}(0, y^2)}
\end{align}
with the numerator taken from the data and the denominator from the fit of
\eqref{K-def}.

%%%%%%%%%%%%%%%%%%%%%%%%%%%%%%%%%%%%%%%%

\subsection{Factorization of \texorpdfstring{$I(\zeta,y^2)$}{I(zeta, y2)}}
\label{sec:factorization}

We assume that in the $y$ range specified by \eqref{y-selected-range}
the $\zeta$ and $y$ dependence of $I(\zeta,y^2)$ factorizes as
\begin{equation}
   \label{product-ansatz}
   I_{q_1 q_2}(\zeta,y^2)
   =
   J_{q_1 q_2}(\zeta) \,
   K_{q_1 q_2}(y^2)
   \,,
\end{equation}
where $K_{q_1 q_2}$ is given in \eqref{K-def} and $J_{q_1 q_2}$ is an even
function of $\zeta$.  Fourier transforming \eqref{product-ansatz} w.r.t.\
$\zeta$ and comparing with \eqref{inv-fct-mellin}, we obtain
the relation
\begin{align}
   \label{J-from-Ahat}
   \frac{1}{2\pi} \int_{-1}^{1} \dd \zeta\,
      e^{i \zeta \omega} \, J_{q_1 q_2}(\zeta)
   &=
   \hat{A}_{q_1 q_2}(\omega)
   \,,
\end{align}
where $\hat{A}_{q_1 q_2}$ defined in \eqref{A-hat-def} is now $y$ independent.
This implies the normalization condition
\begin{align}
   \label{normalize-J}
   \int_0^1 \dd \zeta\, J_{q_1 q_2}(\zeta)
   &= \pi
   \,.
\end{align}
The different models used for the functional dependence of $J_{q_1 q_2}$ are
specified below.

We will validate the factorization hypothesis \eqref{product-ansatz} by
performing fits to \eqref{J-from-Ahat} in narrow bins of $y$ and monitoring to
which extent the functions $J_{q_1 q_2}$ obtained in this way depend on $y$.
Details are given in Section~\ref{sec:local-fits}.

%%%%%%%%%%%%%%%%%%%%%%%%%%%%%%%%%%%%%%%%

\subsection{Dependence on the distance \texorpdfstring{$y$}{y}}
\label{sec:y-dependence}

\begin{figure}
\begin{center}
\subfloat[\label{fig:y-dep-data}]{
\includegraphics[height=0.38\textwidth,trim=10 0 5 0,clip]{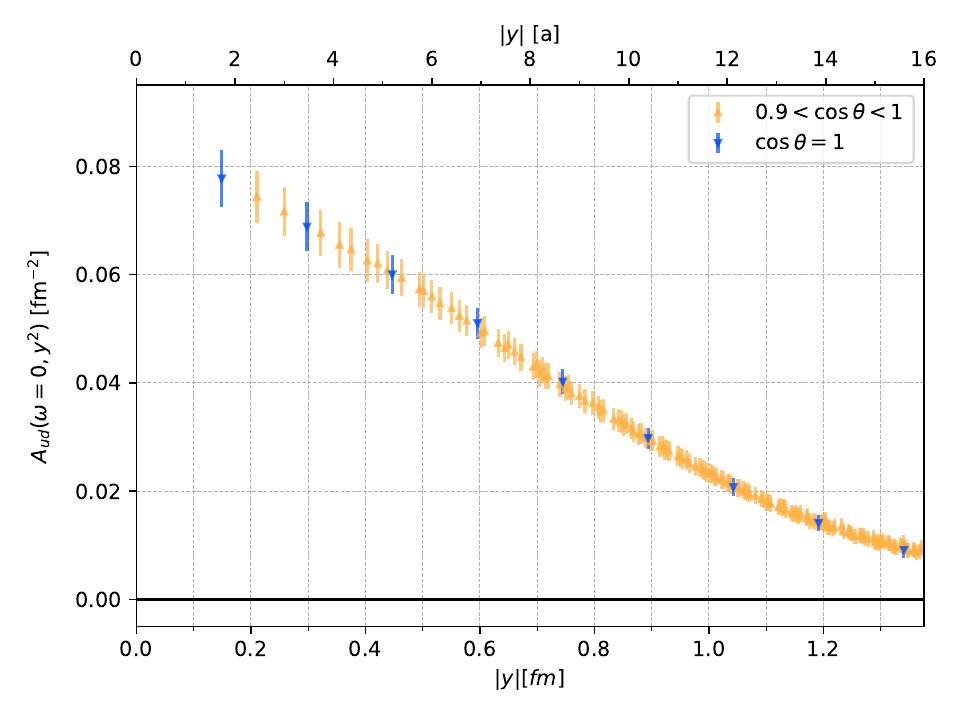}
}
\subfloat[\label{fig:y-dep-fit} ]{
\includegraphics[height=0.38\textwidth,trim=10 0 5 0,clip]{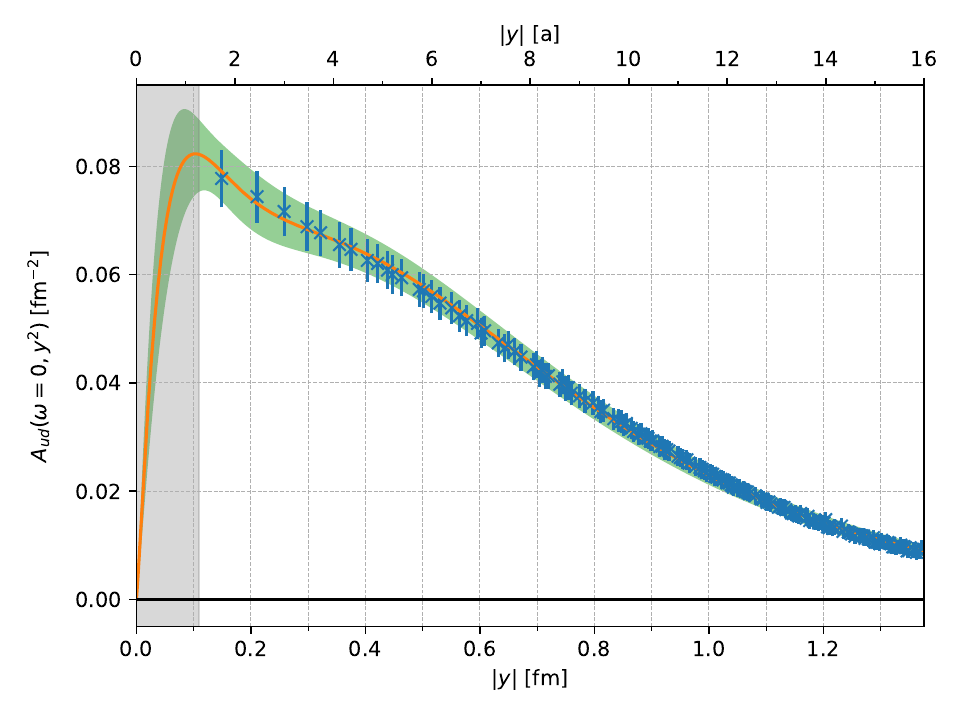}
}
\caption{\label{fig:y-dependence} Lattice data of the invariant function $A_{u
d}$ for $\omega = 0$, together with the fit \protect\eqref{y-ansatz}.  The
shaded region $y \le 0.11 \fm$ is not used in our analysis.}
\end{center}
\end{figure}

From now on we focus on the quark flavor combination $(q_1, q_2) = (u d)$ for
the reasons given below Equation~\eqref{eq:phys_me_decomp}.
\rev{Our data for $A_{u d}(\omega = 0, y^2) = K_{u d}(y^2)$ is quite precise.
Figure~{\ref{fig:y-dep-data}} shows that there is no systematic difference
between data points with $0.9 < \cos\theta < 1$ and data points with $\cos\theta
= 1$. This suggests that within the range $0.9 < \cos\theta$ of our data
selection, possible anisotropy effects are smaller than statistical errors for
this observable.}

For the function $K_{u d}(y^2)$ we adapt the fit previously performed
in~\cite{Bali:2021gel}, which has the form
\begin{align}
   \label{y-ansatz}
   K_{u d}(y^2)
   &=
   A_1 \, (\eta_1 \ms y)^\delta \, e^{-\eta_1(y - 4 a)}
   + A_2 \, (\eta_2 \ms y)^\delta \, e^{-\eta_2(y - 4 a)}
   &
   \text{ for } \sqrt{3} \ms a \le y \le 16 a
\end{align}
with $\delta = 1.2$.  The fitted parameters $A_i$ and $\eta_i$ are
\begin{align}
   \label{y-fit}
   A_1 &= (3.49 \pm 2.37) \times 10^{-4} \ms \fm^{-2}
   \,,
   &
   A_2 &= (5.14 \pm 0.04) \times 10^{-2} \ms \fm^{-2}
   \,,
   \notag \\
   \eta_1 &= (17.6 \pm 2.8) \fm^{-1}
   \,,
   &
   \eta_2 &= (3.53 \pm 0.03) \fm^{-1}
   \,.
\end{align}
In Figure~\ref{fig:y-dep-fit} we show the fit together with the data.  We
emphasize that the $y$ region where the function \eqref{y-ansatz} increases
with $y$ due to the factor $y^\delta$ is never used in our analysis.

At this stage we include data below $y = 4 a$ for the following reason.  In
order to evaluate the number sum rule \eqref{eq:sr-proton}, we need values of
$y$ down to $0.11 \fm \approx 1.29 \ms a$.  Fitting the form \eqref{y-ansatz}
down to $y = 4 a$ would entail a substantial extrapolation error at smaller $y$.
 By contrast, a fit down to $y = \sqrt{3} \ms a \approx 1.73 \ms a$, which
corresponds to the lattice points closest to the origin along the space
diagonals, requires only moderate extrapolation.

Together with the factorization assumption \eqref{product-ansatz}, our fit of
$K_{u d}$ relates $J_{u d}$ with the number sum rule as
\begin{align}
   \label{sr-from-J}
   S_{u d}(y_{\text{cut}})
   &=
   f(y_{\text{cut}}) \, J_{u d}(\zeta = 0)
\intertext{with}
   \label{sr-factor}
   f(y_{\text{cut}})
   &= 2\pi \int_{y_{\text{cut}}}^\infty \dd y\, y \, K_{u d}(y^2)
   \approx
   0.215
   \,,
\end{align}
where we take $y_{\text{cut}} = 0.11 \fm$ from \eqref{y-cut}.

The proportionality factor $f$ decreases by about $16 \%$ if we integrate over
$y$ up to the upper limit $16 \ms a$ of our fit, rather than up to $\infty$. In
turn, it decreases by about $11 \%$ if we integrate over $y$ from $4 a$ to
$\infty$.  These numbers may be taken as upper limits (or very conservative
estimates) for the uncertainty on $f$ due to $(i)$ extrapolating the fit
\eqref{y-ansatz} to $y \ge 16 \ms a$, and due to $(ii)$ including the region $y
\le 4 a$ where discretization effects in our data may be substantial.
We also recall from Section~\ref{sec:sumrule} that the value of $y_{\text{cut}}$
in the sum rule integral itself is a matter of choice.  Multiplying or dividing
$y_{\text{cut}}$ by $2$ changes the $f$ by less than $5 \%$.

%%%%%%%%%%%%%%%%%%%%%%%%%%%%%%%%%%%%%%%%

\subsection{Dependence on the skewness \texorpdfstring{$\zeta$}{zeta}}
\label{sec:zeta-dependence}

The main purpose of this work is to explore different forms for the dependence
of Mellin moments on $\zeta$.  We now present our models for this dependence.

For ease of notation, we assume that $|\zeta| \le 1$ throughout this section,
bearing in mind that $J_{q_1 q_2}(\zeta)$ is zero for $|\zeta| > 1$.  The
Fourier transform of $J_{q_1 q_2}$ is hence to be understood as in
\eqref{inv-fct-mellin}.  We will also omit the quark flavor labels henceforth,
given that we only perform fits for $(q_1 q_2) = (u d)$ as explained above.

%%%%%%%%%%%%%%%%%%%%%%%%%%%%%%%%%%%%%%%%

\paragraph{The polynomial model} is what we used in our previous work
\cite{Bali:2021gel}.  For a given integer $N$ it is given by
\begin{align}
\label{poly-J}
J_{\mathrm{poly}}\left(\zeta\right)
&=
\pi \sum_{n, m=0}^N \zeta^{2 n} \left(T^{-1}\right){\!}_{n m} \, c_{m}
&
\text{ with } c_0 = 1
\,,
\end{align}
where the $(N+1) \times (N+1)$ matrix $T$ is defined by
\begin{equation}
   T_{m n}
   =
   1 \big/ (1+2 n+2 m)
   \,.
\end{equation}
The condition $c_0 = 1$ ensures the normalization condition \eqref{normalize-J},
as shown in Section~4.4 of~\cite{Bali:2021gel}.  The model thus has $N$ free
parameters, $c_1$ to $c_N$.
The normalized Fourier transform of \eqref{poly-J} reads
\begin{equation}
\hat{A}_{\mathrm{poly}}\left(\omega\right)
=
\sum_{n, m=0}^N h_n(\omega) \, \left(T^{-1}\right){\!}_{n m} \, c_{m}
\,,
\end{equation}
with the functions $h_n$ given by
\begin{equation}
\label{poly-h}
h_n(\omega)
=
\frac{1}{2} \int_{-1}^{1} \dd \zeta \, e^{i \zeta \omega} \, \zeta^{2 n}
=
\sin (\omega) \ms s_n(\omega) + \cos (\omega) \ms c_n(\omega)
\end{equation}
and
\begin{equation}
s_n(\omega)
=
\sum_{m=0}^n \frac{(-1)^m \, (2 n)!}{(2 n-2 m)!} \,
   \frac{1}{\omega^{1+2 m}}
\,,
\quad
c_n(\omega)
=
\sum_{m=0}^{n-1} \frac{(-1)^m \, (2 n)!}{(2 n-2 m-1)!} \,
   \frac{1}{\omega^{2+2 m}}
\,.
\end{equation}
We note that the singular behavior of $s_n(\omega)$ and $c_n(\omega)$ for
$\omega \to 0$ cancels in the combination \eqref{poly-h}, such that the function
$h(\omega)$ is finite at $\omega = 0$, as is obvious from its integral
representation.

In \cite{Bali:2021gel} we carried out fits with $N = 2$ and $N = 3$ and found
their results to be quite close to each other.  No clear improvement  was
observed in the description of the data for $N = 3$, and we regard that version
as close to overfitting the data.  In the present work we hence restrict
ourselves to the case $N = 2$.

The polynomial model does not satisfy the properties we derived in
Section~\ref{sec:moment-properties}: the form \eqref{poly-J} does not vanish for
$\zeta \to 1$ (unless one introduces an additional constraint on the polynomial
coefficients), nor does it allow for a non-analytic behavior at $\zeta = 0$ or
$\zeta = 1$.

\begin{figure}
\begin{center}
\subfloat[]{
\includegraphics[height=0.22\textwidth]{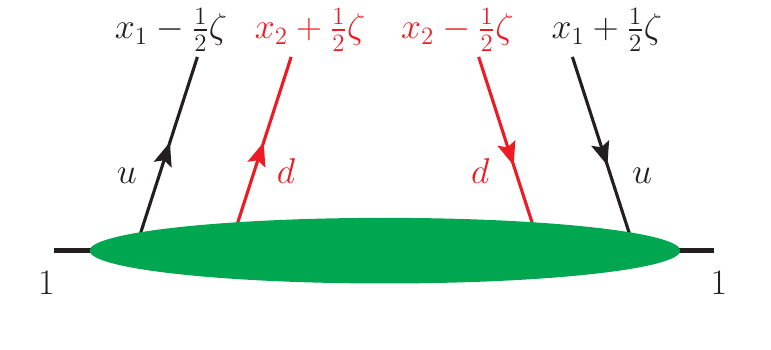}
}
\hspace{2em}
\subfloat[]{
\includegraphics[height=0.22\textwidth]{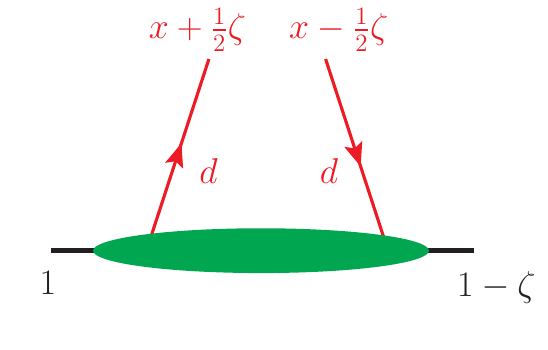}
}
\caption{Comparison between a skewed DPD $F_{u d}\left(x_i, \zeta, y^2\right)$
in the region $x_i \pm \zeta / 2 > 0$ (a) and a generalized (a.k.a.\ skewed)
single-parton distribution (b).  The notation in
\protect\cite{Radyushkin:1997ki} uses the variable $X = x + \half\zeta$ instead
of $x$.  All momentum fractions refer to the incoming proton.}
\label{fig:dpd-gpd}
\end{center}
\end{figure}

When making this ansatz in our previous work, we had in mind the case
of generalized single-parton distributions (GPDs).\footnote{%
The following statements hold for both for the skewness parameter $\zeta$ of
Radyushkin  \protect\cite{Radyushkin:1997ki} and for the parameter $\xi$ of Ji
\protect\cite{Ji:1998pc}, given their relation $\xi = \zeta / (2 - \zeta)$.}
Their Mellin moments \emph{are} in fact polynomials in the skewness parameter
$\zeta$, as a consequence of Lorentz invariance \cite{Ji:1998pc}.  Moreover,
they do not generally vanish for $\zeta \to 1$, which already follows from the
fact that the lowest moments of quark GPDs are $\zeta$ independent.  We note
that the limit $\zeta \to 1$ is peculiar for GPDs, because for a given invariant
momentum transfer $t$ it is always outside the physical region $-t \ge m_p^2 \,
\zeta^2 / (1 - \zeta)$ in which the proton states are on shell.
Despite the apparent similarity between the role of $\zeta$ in DPDs and GPDs
(see Figure~\ref{fig:dpd-gpd}), the $\zeta$ dependence of these distributions
and their Mellin moments is therefore very different.

The following three models are new and have been designed to fulfill the
properties discussed in Section~\ref{sec:moment-properties}, i.e.\ they ensure
$I \to 0$ for $\zeta \to 1$, and they allow for a nonanalytic behavior of $I$ at
$\zeta = 0$ and $\zeta = 1$.

%%%%%%%%%%%%%%%%%%%%%%%%%%%%%%%%%%%%%%%%

\paragraph{The power-law model} is a rather direct generalization of the polynomial
model, given by
\begin{align}
   \label{pow-J}
   J_{\mathrm{pow}}(\zeta)
   &=
   \pi (a+1) \, \bigl( 1 - |\zeta| \bigr)^a
   &
   \text{ with } a > 0
   \,.
\end{align}
Its normalized Fourier transform reads
\begin{equation}
\hat{A}_{\mathrm{pow}}(\omega)
=
{_1}F_2\left(1;\frac{a+2}{2},\frac{a+3}{2}; - \frac{\omega^2}{4}\right)
\,,
\end{equation}
where ${_1}F_2$ is the hypergeometric function.

The function \eqref{pow-J} has a spike (i.e.~a discontinuous first derivative) at
$\zeta=0$ for all admissible values of $a$.

%%%%%%%%%%%%%%%%%%%%%%%%%%%%%%%%%%%%%%%%

\paragraph{The integral model} has the form
\begin{align}
   \label{int-J}
   J_{\mathrm{int}}(\zeta)
   &=
   \pi \, \frac{\Gamma(a+b+1)}{\Gamma(a+1)\Gamma(b)}
   \int_{|\zeta|}^1 \dd z \, z^{a-1} (1-z)^{b-1}
   &
   \text{ with } a > 0 \text{ and } b > 0
   \,,
\end{align}
and its normalized Fourier transform is given by
\begin{align}
\hat{A}_{\mathrm{int}}(\omega)
&=
{_2}F_3\left(\frac{a+2}{2}, \frac{a+1}{2};
   \frac{3}{2}, \frac{a+b+2}{2}, \frac{a+b+1}{2}; - \frac{\omega^2}{4}\right)
\end{align}
with the generalized hypergeometric function ${_2}F_3$.  The ansatz
\eqref{int-J} is inspired by the leading-order splitting formula
\eqref{splitting-lowest-moment} for $g\rightarrow q\bar{q}$ in conjunction with
a PDF behaving as in \eqref{gluon-PDF}, without trying to emulate its detailed
features, which are specific to $I_{q q}$ and not directly applicable to $I_{u
d}$.

For $a \le 1$ the function \eqref{int-J} has a spike at $\zeta = 0$, whereas for
$a > 1$ it is smooth at that point (with a vanishing first derivative but
possible discontinuities of higher odd derivatives).

%%%%%%%%%%%%%%%%%%%%%%%%%%%%%%%%%%%%%%%%

\paragraph{The cosine model} is inspired by the shape of the integral model, but
allows for a more direct interpretation of the parameters. It is given by
\begin{align}
   \label{cos-J}
   J_{\mathrm{cos}}(\zeta)
   =
   \mathcal{N}_{a,b} \, \cos^{\ms b} \left( \frac{\pi}{2} \, |\zeta|^a \right)
   &
   \text{ with } a > 0 \text{ and } b > 0
   \,,
\end{align}
where $\mathcal{N}_{a, b}$ is determined by the normalization
condition~\eqref{normalize-J}.
We have not found a closed form for its Fourier transform, which is computed
numerically in our fit to the lattice data.
Important features of the function $J_{\mathrm{cos}}(\zeta)$ are:
\begin{itemize}
\item the parameter $a$ controls the width of the central maximum around $\zeta
= 0$,
\item the parameter $b$ controls the speed of approaching zero for $|\zeta| \to
1$,
\item for $a \le 1/2$ it has a spike at $\zeta = 0$ and is very sensitive to even
small changes in $a$, whilst for $a > 1/2$ it is smooth at that point,
\item for positive half integer $a$, the function is analytic at $\zeta=0$.
\end{itemize}

%%%%%%%%%%%%%%%%%%%%%%%%%%%%%%%%%%%%%%%%%%%%%%%%%%%%%%%%%%%%%%%%%%%%%%%%%%%%%%%%

\section{Fit results}
\label{sec:fit-results}

We now discuss the results of fitting the above models to our lattice data.  We
performed global fits to the data in the range
\begin{align}
   \label{fit-range}
   4 a & \le y \le 14 a
   \,,
   &
   - 2.72 \, y \big/ (4 a) & \le \omega \le 2.72 \, y \big/ (4 a)
   \,,
\end{align}
where the second condition follows from the bound \eqref{omega-bound} and from
the range of proton momenta in our simulations.  The highest value of $|\omega|$
in the data thus selected is about $9.5$.  Since statistical errors increase
with the proton momentum \emph{and} with the distance $y$ between the currents,
the uncertainty on the data is smallest around $\omega = 0$ and grows rather
quickly with $|\omega|$.  We omit distances $y > 14 a$ from the fit because some
of the corresponding data points have very large errors.

\begin{figure}
\begin{center}
\includegraphics[width=0.65\textwidth]{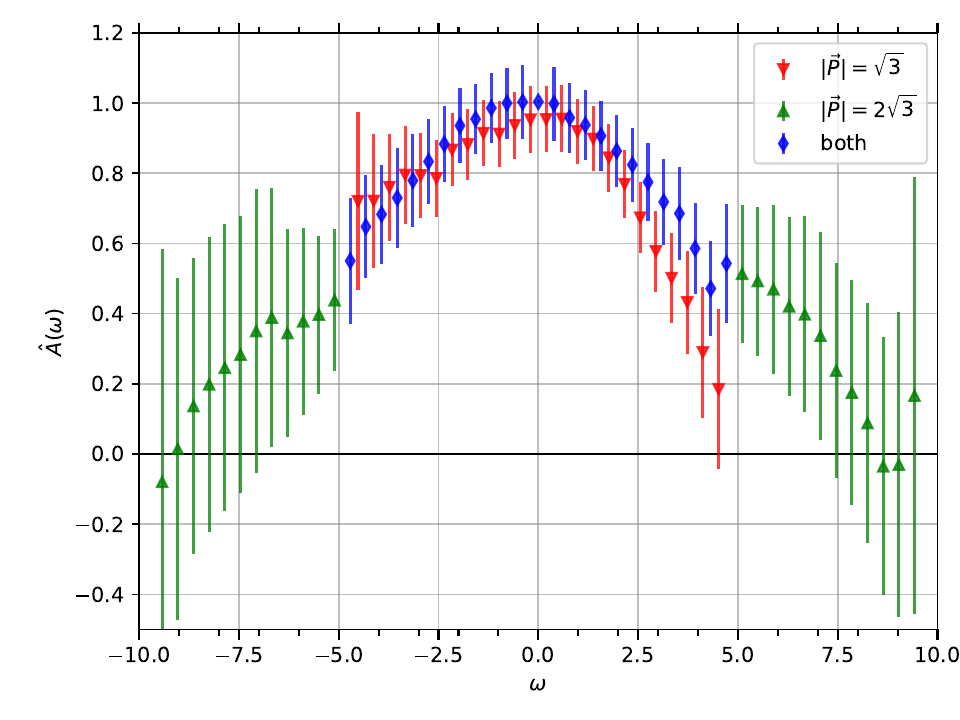}
\caption{\label{fig:omega-dependence} \rev{Data for the normalized invariant
function $\hat{A}(\omega)$, obtained as described in the text.  The different
symbols and colors indicate whether a data point receives contributions from
proton momenta with $|P| = \sqrt{3}$, $|P| = 2 \sqrt{3}$, or from both.}}
\end{center}
\end{figure}

The fitted data are shown in Figure~\ref{fig:omega-dependence}.  They are
obtained by dividing the data for $A(\omega, y^2)$ in the range
\eqref{fit-range} by the function $K(y^2)$, which is evaluated with the fit
specified in \eqref{y-ansatz} and \eqref{y-fit}.  The errors on the data points
are obtained with the jackknife method.

\rev{Before turning to our fits, let us discuss some aspects of the data. There
are many cases in which several pairs of $\mvec{y}$ and $\mvec{p}$ have the same
$\omega$ and $y^2$, even with our cut \eqref{theta-cut}.  Some of these pairs
are related by symmetries that are exact on the lattice (such as permutations or
sign reversal of vector components), but others are not.  For the vector
$\mvec{P} = - (1, 1, 1)$, which is related  to the proton momentum $\mvec{p}$ by
Eq.~\eqref{lattice-momenta}, we obtain for instance the same $\omega$ with
$\mvec{y} = (2,6,7)\ms a$ and $\mvec{y} = (3,4,8)\ms a$, both of which have
length $y = \sqrt{89}\ms a$.}

\rev{If we focus on the $\omega$ dependence, there are even more configurations
that can be combined.   For $\mvec{P} = - (1, 1, 1)$, the same $\omega$ is for
instance obtained with the vectors $\mvec{y}$ just mentioned, with the two
vectors $\mvec{y} = (3,6,6)\ms a$ and $\mvec{y} = (4,4,7)\ms a$ of length $y =
9\ms a$, with 9 further vectors with different values of $y$ between
$9\ms a$ and $10\ms a$, and with yet more vectors with $y$ outside that range.
The same $\omega$ may also be obtained for different proton momenta; a trivial
example is $\mvec{P} = - (1, 1, 1)$ with $\mvec{y} = (6,6,6)\ms a$ and $\mvec{P}
= - (2,2,2)$ with $\mvec{y} = (3,3,3)\ms a$.}

\rev{Each data point in Figure~\ref{fig:omega-dependence} results from combining
all pairs of $\mvec{y}$ and $\mvec{p}$ that give the same $\omega$. Even with
this combination the result is still statistics limited, especially for larger
values of $|\omega|$.  Averaging only over pairs of $\mvec{y}$ and $\mvec{p}$
that are related by lattice symmetries would give points with yet larger
statistical errors and not permit us to assess the size of lattice
artifacts in the $\omega$ dependence.}

\rev{The figure shows that data points receiving only contributions from
$|\mvec{P}| = \sqrt{3}$ tend to deviate systematically from those with
contributions from $|\mvec{P}| = \sqrt{3}$ and $2 \sqrt{3}$.  We recall that we
have only one momentum with $|\mvec{P}| = \sqrt{3}$, whereas the data with
$|\mvec{P}| = 2 \sqrt{3}$ includes four momenta and thus benefits from averaging
over different spatial directions.  We also note that time reversal (which flips
the sign of $\mvec{p}$ but not of $\mvec{y}$) implies that the $\omega$
distribution must be symmetric, whilst the central values of the $|\mvec{P}| =
\sqrt{3}$ data are slightly asymmetric in $\omega$.  We shall not investigate
this issue  further, because the observed deviations are not significant within
statistical errors.}

In Table~\ref{tab:fit-results} we list the results of the fits that will be
discussed in the following.  We note that the $\chi^2$ of the data was computed
\emph{without} taking into account the statistical correlations between data
points at different $\omega$.  This is because the corresponding
correlation matrix is very close to singular, such that its inversion would be
affected by severe instabilities.  With correlations between data points being
neglected, the resulting values of $\chi^2$ per degree of freedom are
significantly below $1$ in all our fits.

\begin{table}
\centering
\begin{tabular}{|c|c|c|c|}
\hline \hline
model & \multicolumn{2}{c|}{parameters} & $\chi^2 /$ d.o.f. \\
\hline \hline
polynomial & $c_1 = (9.84 \pm 4.41) \times 10^{-2}$
           & $c_2 = (6.16 \pm 4.57) \times 10^{-2}$ & 0.38 \\
\hline
power-law  &  $a = 2.95 \pm 1.10$ & --- &  0.36 \\
\hline
integral   & $a = 0.37 \pm 2.83$ & $b = 2.04 \pm 3.97$ & 0.35 \\
integral   & $a=0.5$ fixed & $b = 1.23 \pm 1.84$ &  0.35 \\
integral   & $a=4$\phantom{$.5$} fixed & $b = 7.36 \pm 2.67$ & 0.40 \\
\hline
cosine     & $a = 0.29 \pm 1.14$ & $b = 1.92 \pm 3.33$ & 0.35 \\
cosine     & $a=1$\phantom{$.5$} fixed & $b = 5.42 \pm 1.78$ & 0.38 \\
\hline \hline
\end{tabular}
\caption{Fit results for our models of the $\zeta$ dependence.  The low
values of $\chi^2 /$ d.o.f.\ are explained in the text.  The models are
specified in Section~\protect\ref{sec:zeta-dependence}.}
\label{tab:fit-results}
\end{table}

In the left panels of the following figures, we show the fits of
$\hat{A}(\omega)$ with their error bands, together with the fitted data points.
In the right panels of the same figures, we show the resulting value of
$J(\zeta)$, multiplied by the factor $f$ from \eqref{sr-factor}.  We recall
that, according to the number sum rule \eqref{eq:sr-proton}, one should have $f
J(\zeta = 0) = 2$ up to corrections of parametric order $\alpha_s^2$ or
$(\Lambda \ms y_{\text{cut}})^2$.

%%%%%%%%%%%%%%%%%%%%%%%%%%%%%%%%%%%%%%%%

\subsection{Comparison of models}

\begin{figure}
\begin{center}
\subfloat[polynomial model]{
\includegraphics[width=0.49\textwidth]{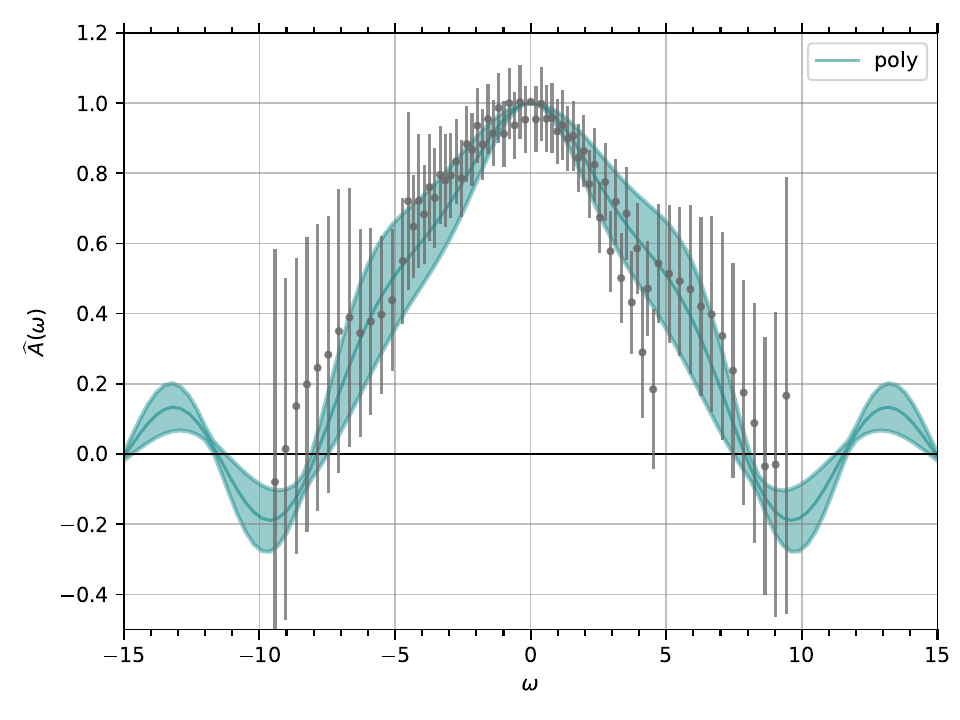}
\includegraphics[width=0.49\textwidth]{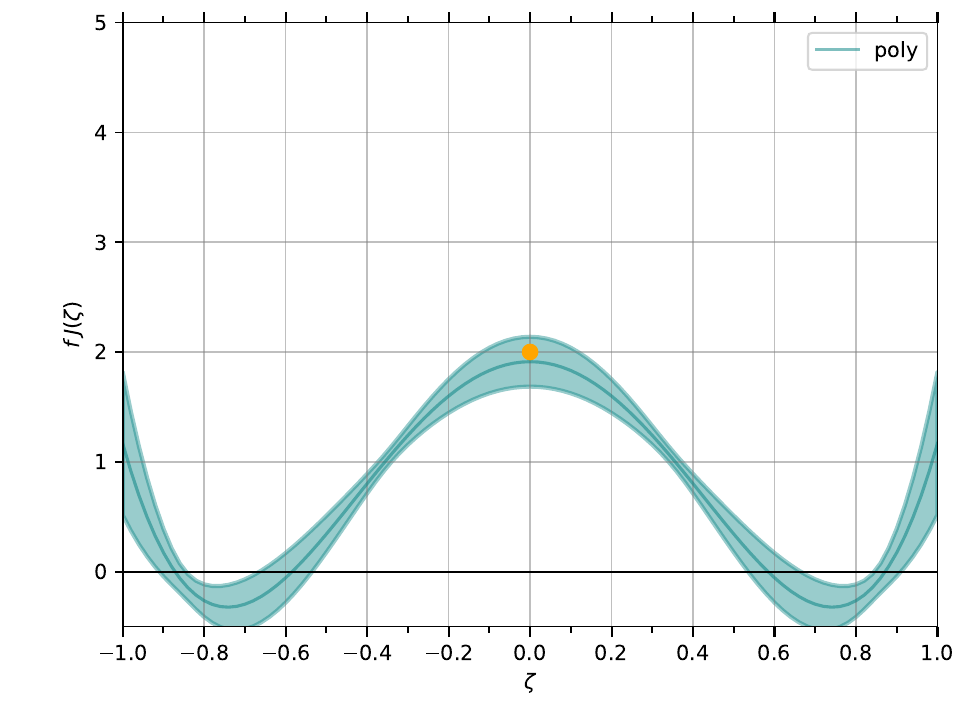}
}
\\[1em]
\subfloat[power-law model]{
\includegraphics[width=0.49\textwidth]{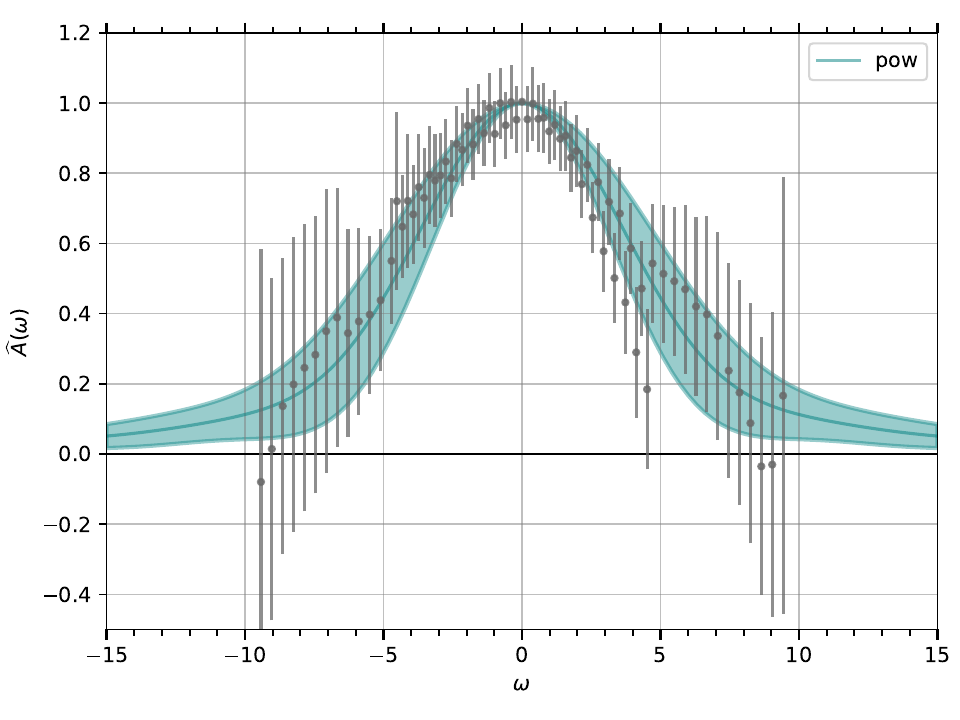}
\includegraphics[width=0.49\textwidth]{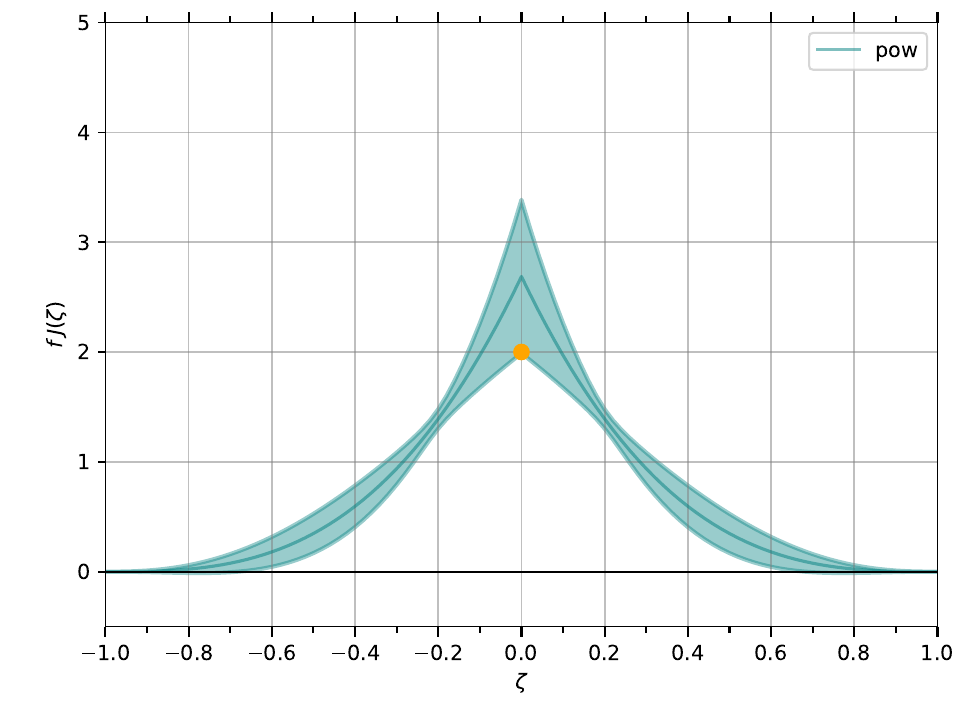}
}
\caption{\label{fig:poly_abs_res} Fit results for the polynomial and  power-law
models.  Left panels show the fit and data vs.\ Ioffe time $\omega$ and right
panels show the fit results after a Fourier transformation to the
skewness~$\zeta$. The plots on the right are normalized such that the number
sum rule requires the value $2$ at $\zeta=0$, which is indicated by an orange
dot. Further explanations are given in the text.}
\end{center}
%\end{figure}
%
\vspace{1em}
%
%\begin{figure}
\begin{center}
\includegraphics[width=0.49\textwidth]{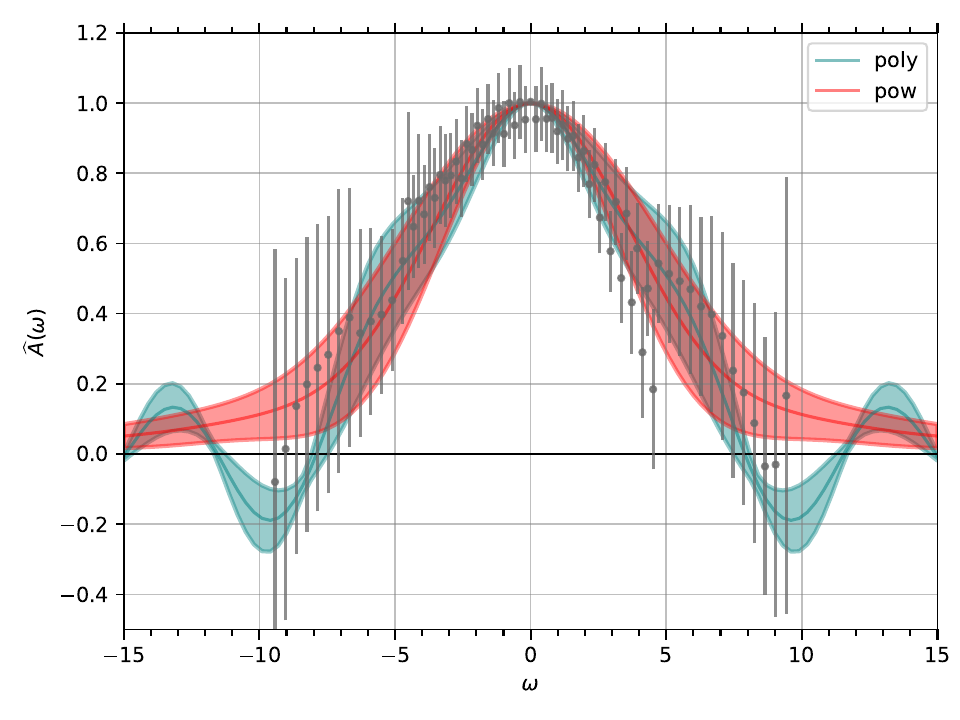}
\includegraphics[width=0.49\textwidth]{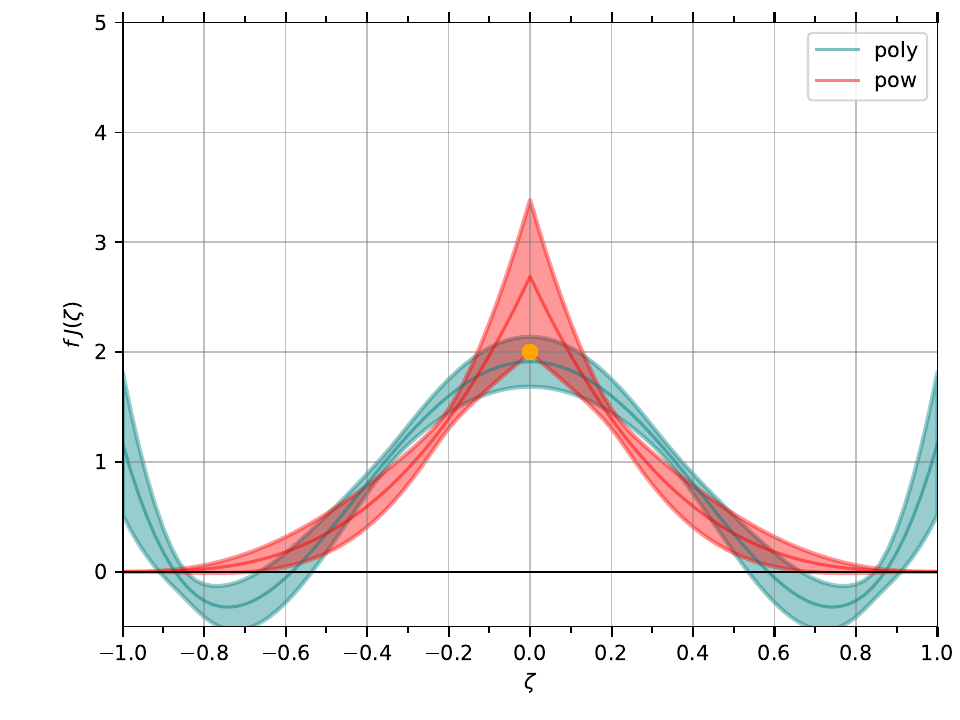}
\caption{\label{fig:phi_poly_abs_comp} Comparison of the fit results shown in
Figure~\protect\ref{fig:poly_abs_res}.  The polynomial model is shown in blue
and the power-law model in red.}
\end{center}
\end{figure}

The fit results for the polynomial and the power-law model are shown in
Figure~\ref{fig:poly_abs_res}.  Both fits give a good description of the data.
We note that at large $|\omega|$ they both yield rather narrow error bands
compared with the errors on the data points.  We observe that the two models
behave rather differently in the $|\omega|$ region where we have no data (or
data with huge errors), with pronounced oscillations occurring in the polynomial
model.\footnote{%
Such oscillations could be anticipated from the functional form in
\protect\eqref{poly-h}.}
In the Fourier conjugate space, these differences are reflected by pronounced
differences between the fit results around $\zeta = 0$.  Whereas the polynomial
model is smooth at that point by construction, the power-law model exhibits a
clear spike.  At $|\zeta| \to 1$ the polynomial model results in a nonzero value
of the Mellin moment, at least within $1\sigma$ uncertainties. Based on our
arguments in Section~\ref{sec:moment-properties}, we hence disfavor this fitting
ansatz on physical grounds.

Figure~\ref{fig:phi_poly_abs_comp} shows that the two fits agree within their
uncertainties for $\hat{A}$ up to $|\omega| \sim 7$.  They yield similar results
for the Mellin moment at intermediate $|\zeta|$ but differ qualitatively for
$|\zeta|$ close to $0$ or to $1$.  Already at this stage we can conclude that in
$\zeta$ space the uncertainty bands of an individual fit can significantly
misrepresent the actual uncertainty on the Mellin moment.  At $\zeta = 0$, both
fits are consistent with the sum rule value of $2$, although for the power-law
model this value is at the lower end of the uncertainty band.

It is interesting to note that, broadly speaking, the two fits have error bands
of similar size, both in $\omega$ and in $\zeta$ space, despite the fact that
the polynomial model has two and the power-law model one free parameter.  This
shows that the amount of flexibility of the functional form is crucial for the
behavior of the fit, at least with data of the quality we are working with.

%%%%%%%%%%%%%%%%%%%%%%%%%%%%%%%%%%%%%%%%

\begin{figure}
\begin{center}
\subfloat[integral model]{
\includegraphics[width=0.49\textwidth]{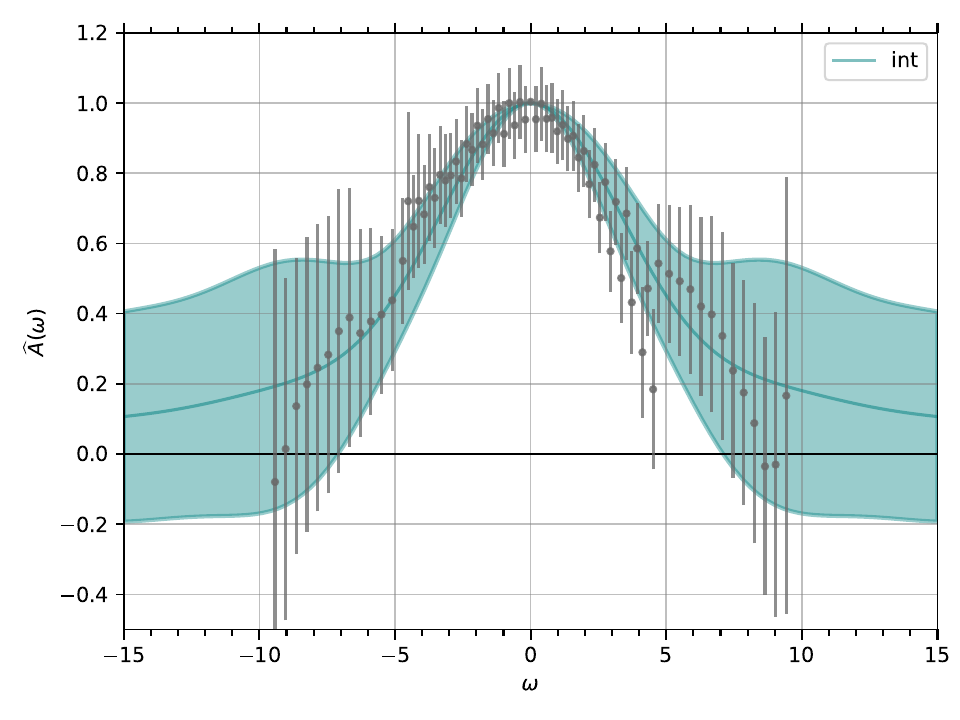}
\includegraphics[width=0.49\textwidth]{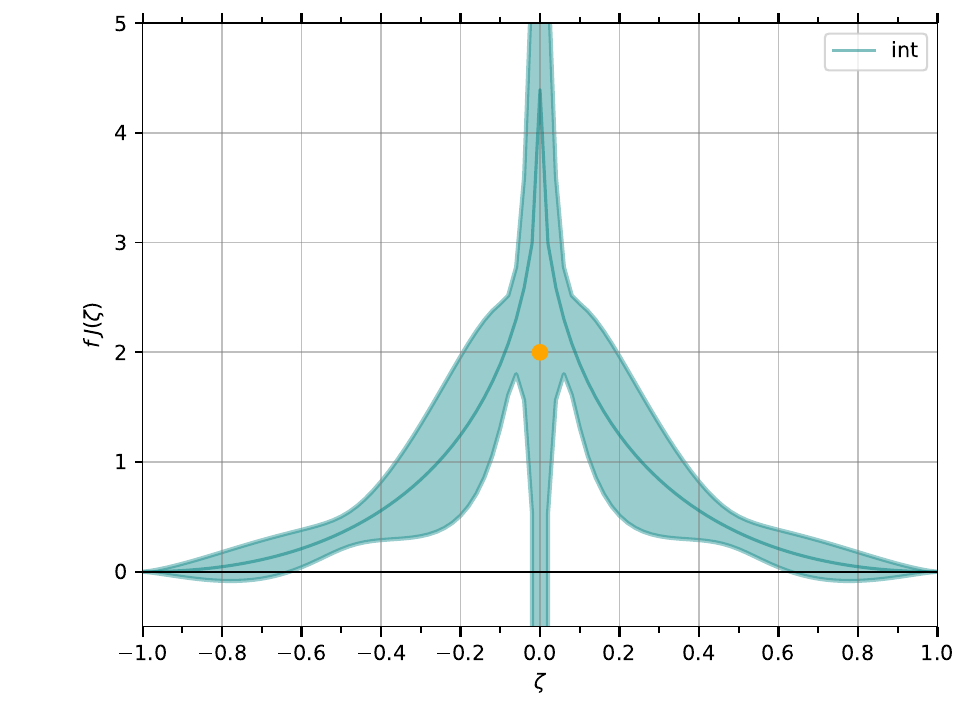}
}
\\[1em]
\subfloat[cosine model]{
\includegraphics[width=0.49\textwidth]{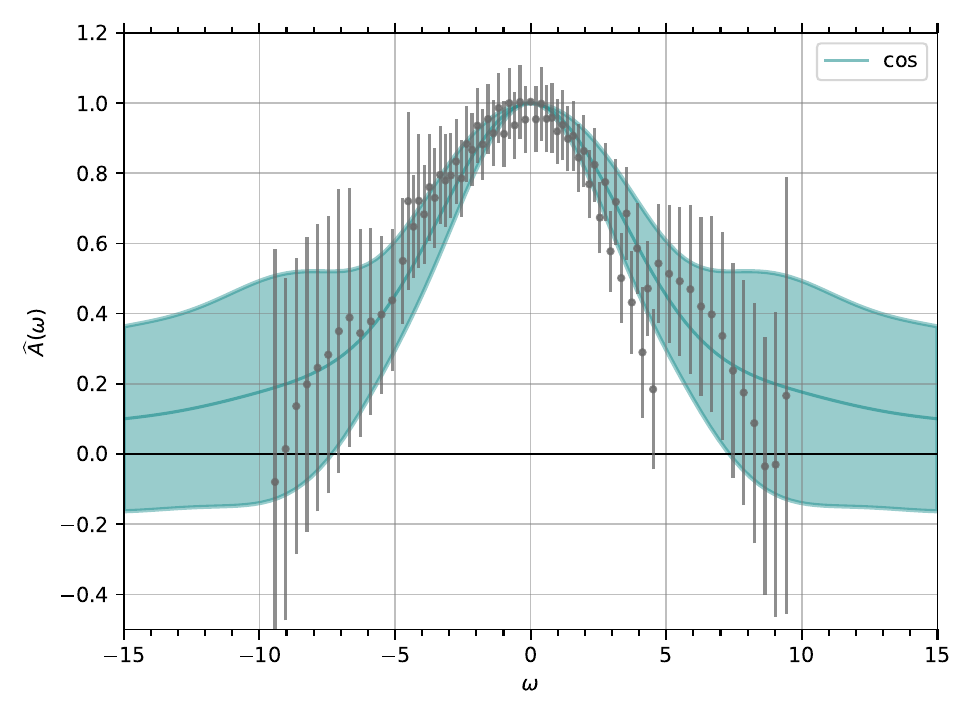}
\includegraphics[width=0.49\textwidth]{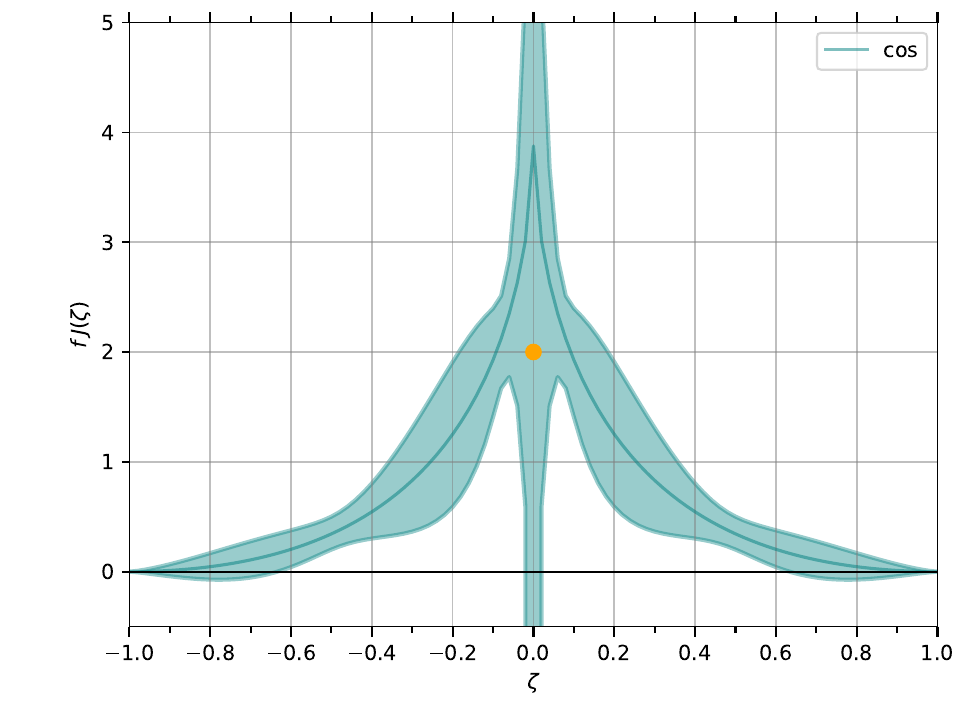}
}
\caption{\label{fig:phi_cosine_res} As Figure~\protect\ref{fig:poly_abs_res},
but for the integral and cosine models.}
\end{center}
% \end{figure}

\vspace{2em}

% \begin{figure}
\begin{center}
\includegraphics[width=0.49\textwidth]
%{A_hat_omega_compdoublepowcosfit_singlephicorr_free_doublepowcosfit_.pdf}
{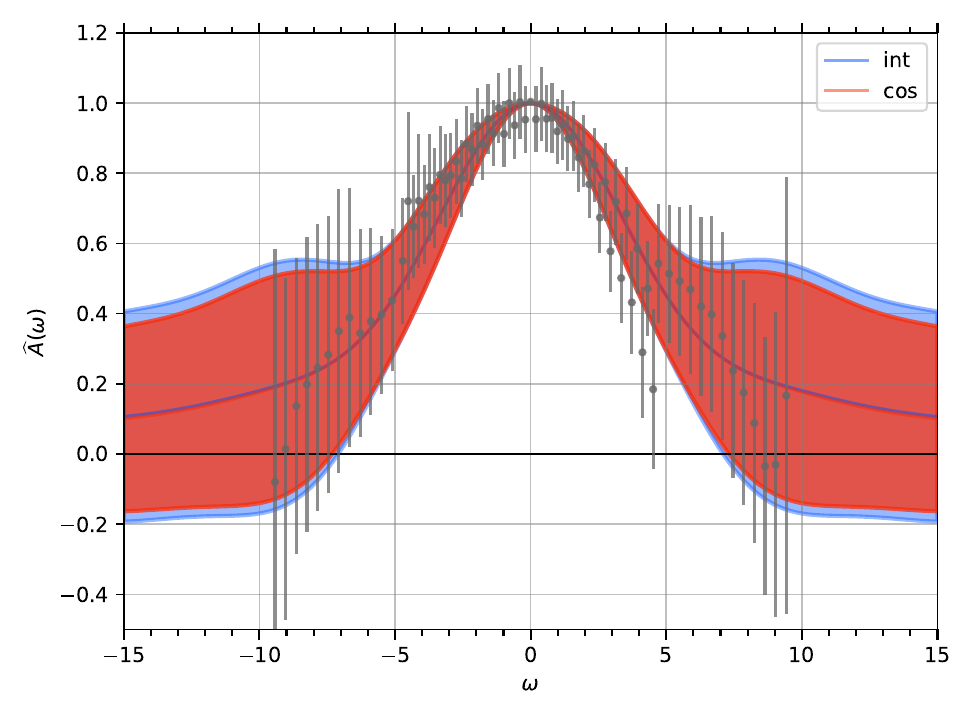}
\includegraphics[width=0.49\textwidth]%{J_zeta_compdoublepowcosfit_singlephicorr_free_doublepowcosfit_.pdf}
{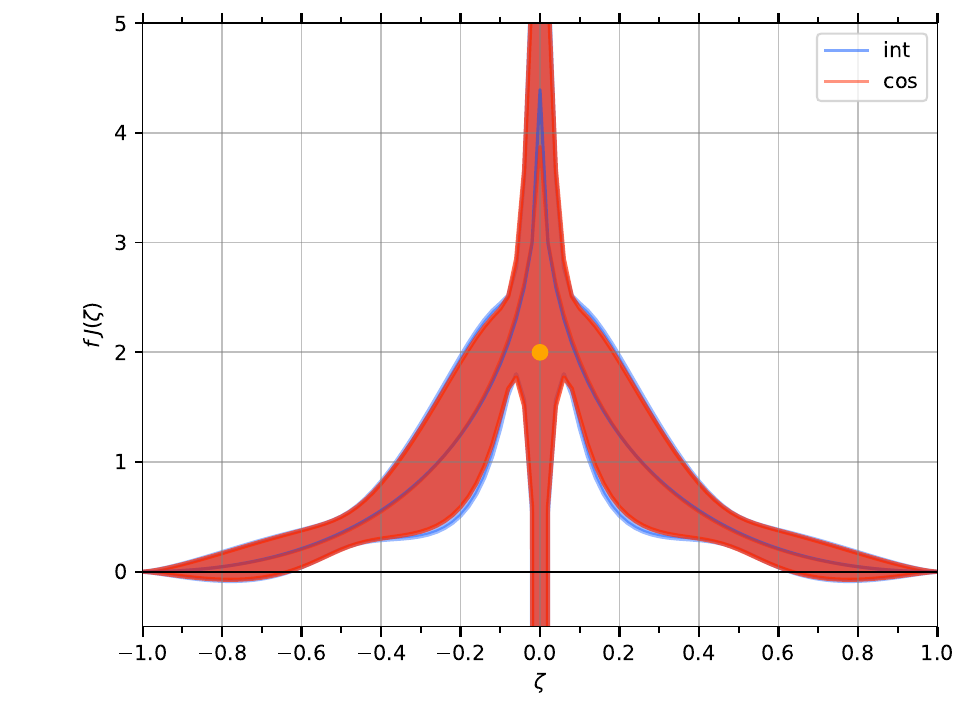}
\caption{\label{fig:phi_cosine_comp} Comparison of the fit results shown in Figure~\protect\ref{fig:phi_cosine_res}.  \rev{The blue and red curves for the central values are barely distinguishable.}}
\end{center}
\end{figure}

The fit results for the integral and cosine models are shown separately in
Figure~\ref{fig:phi_cosine_res} and together in
Figure~\ref{fig:phi_cosine_comp}.  Quite remarkably, both the central values and
the error bands agree almost perfectly between the two fits, despite the
apparent difference between the functional forms \eqref{int-J} and
\eqref{cos-J}.

The fits describe the data well.  Their error bands become broad at large
$|\omega|$, at variance with the polynomial and power-law fits. Correspondingly,
we find a  \emph{huge} uncertainty on the Mellin moment around $\zeta=0$ in the
right panels of Figure~\ref{fig:phi_cosine_res}.  If we accept that these fits
adequately represent the uncertainty of our lattice data at large $|\omega|$, we
must conclude that the data does not permit a reconstruction of the Mellin
moments at the point $\zeta=0$ without further theory input.  Interestingly, the
fit uncertainty on the Mellin moment is extremely large only in a narrow range
$|\zeta| \lsim 0.05$, whereas for larger skewness it is of moderate size.

%%%%%%%%%%%%%%%%%%%%%%%%%%%%%%%%%%%%%%%%

\paragraph{Constrained fits.}

\begin{figure}
\begin{center}
\subfloat[\label{fig:fit_restr_sr} integral and cosine models with fixed $a$]{
\includegraphics[width=0.49\textwidth]{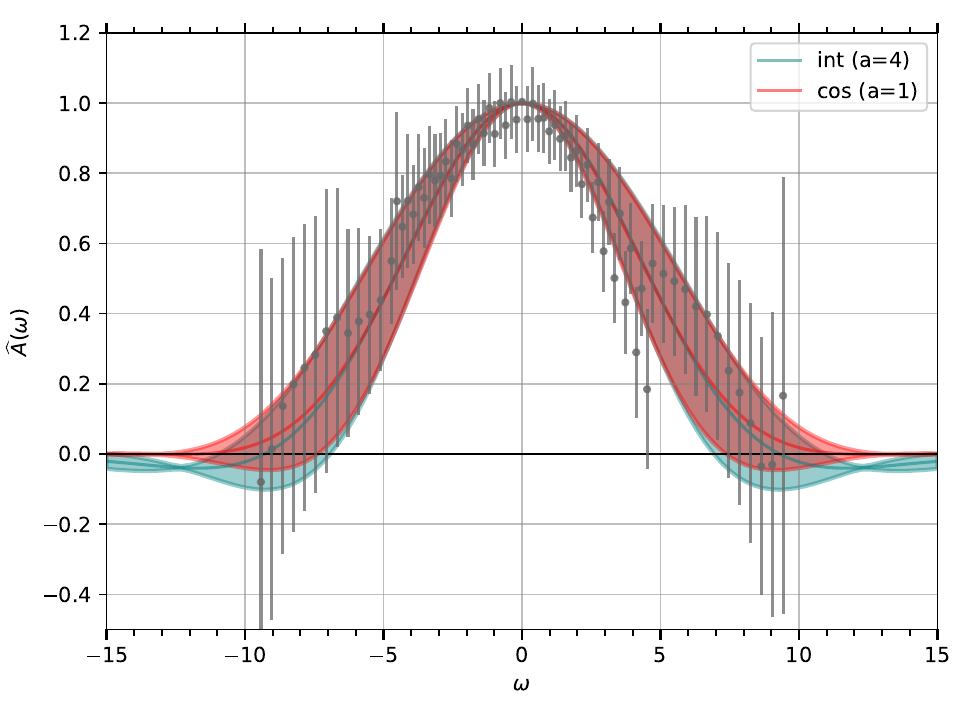}
\includegraphics[width=0.49\textwidth]{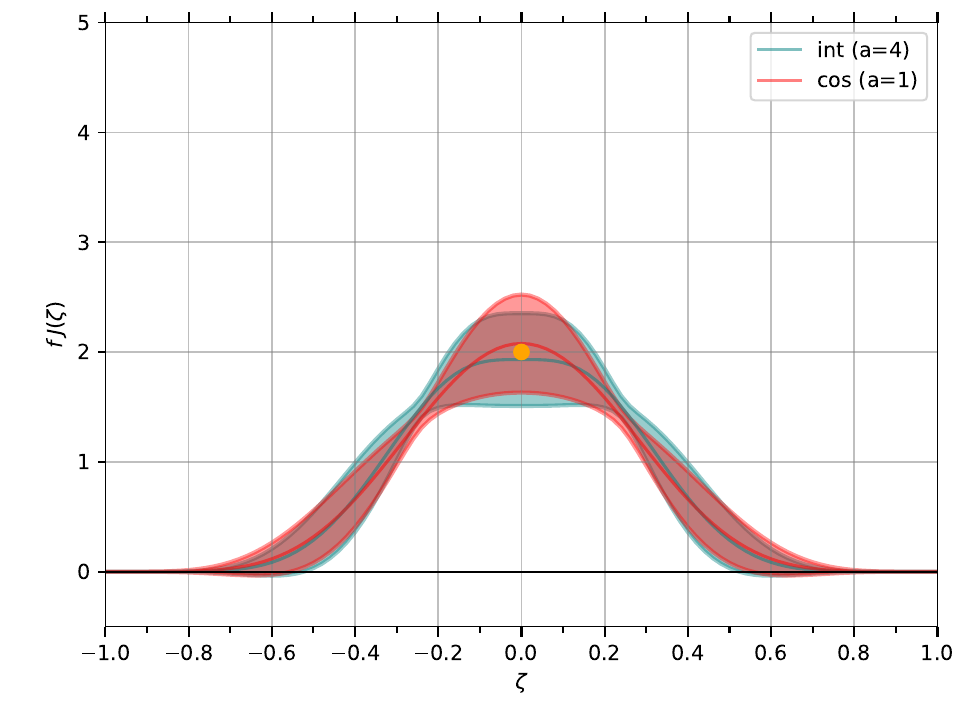}
}
\\[1em]
\subfloat[\label{fig:fit_restr_int} integral model with fixed $a$]{
\includegraphics[width=0.49\textwidth]{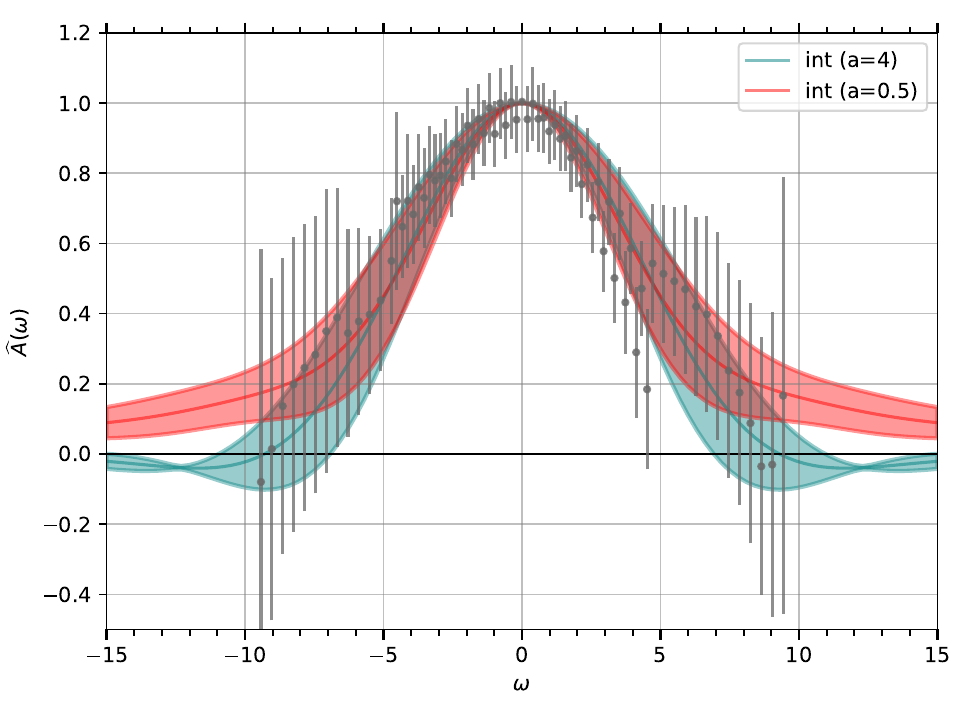}
\includegraphics[width=0.49\textwidth]{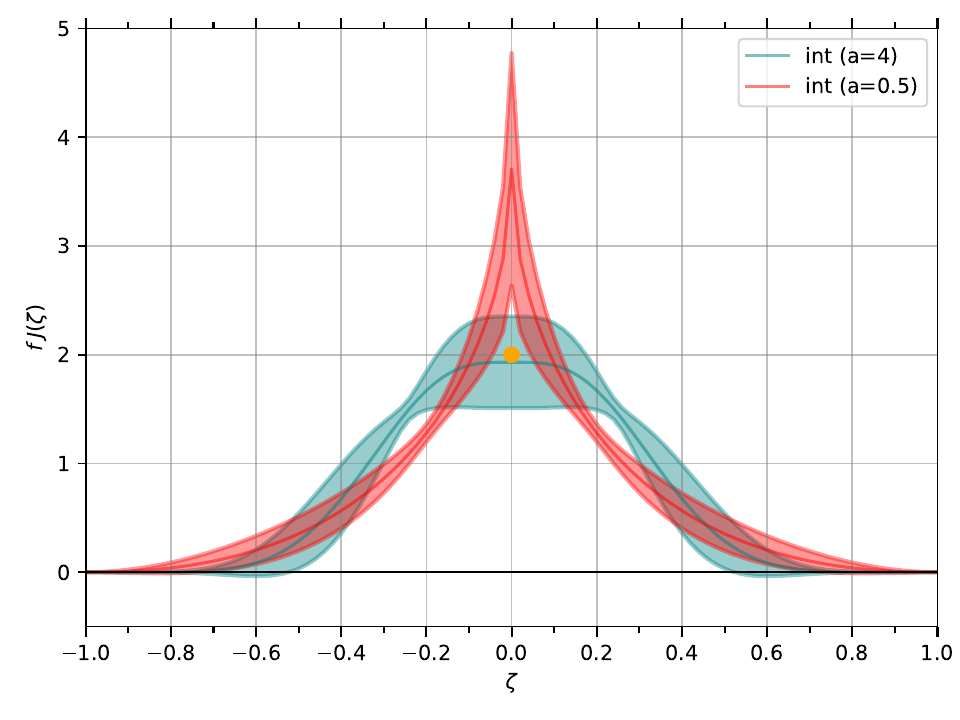}
}
\caption{\label{fig:restricted_fits} As Figure~\protect\ref{fig:phi_cosine_res},
but for fits with one fixed and one free parameter.}
\end{center}
\end{figure}

For the flavor combination $u d$ (but not for equal quark flavors) one may opt
to \emph{use} the number sum rule \eqref{eq:sr-proton} as a theory input to the
fit.  To explore this option, we performed a series of fits to the integral and
cosine models with their respective parameters $a$ fixed to a prior value. We
find that the fit results for $f J(\zeta = 0)$ are close to the sum rule value
$2$ if we impose $a = 4$ in the integral model \eqref{int-J} and $a = 1$ in the
cosine model \eqref{cos-J}. In Table~\ref{tab:fit-results} we see that both
constrained fits have a slightly higher $\chi^2$ than their two-parameter
counterparts, but still provide a good description of the data.  This is also
seen in Figure~\ref{fig:fit_restr_sr}, which additionally shows that the two
constrained fits are quite similar in their central values and error bands.  The
bands are significantly reduced with respect to the two-parameter fits, as is
evident from the comparison with Figure~\ref{fig:phi_cosine_comp}.

For the integral model we also performed a constrained fit with fixed $a = 0.5$,
which is close to the fitted value of $a$ in the two-parameter fit.  The result
is compared with the case $a = 4$ in Figure~\ref{fig:fit_restr_int}.  We see
that the two fits differ noticeably at large $|\omega|$ and result in a quite
different behavior of the Mellin moment for central values of $\zeta$.  The fit
with $a = 0.5$ has a spike at $\zeta = 0$ and misses the sum rule value within
its $1\sigma$ uncertainties.

\begin{table}
\centering
\begin{tabular}{|c|c|c|c|c|}
\hline \hline
model & fixed params. & $S_{u d} = f J_{u d}(\zeta=0)$
      & $100 \ms \langle \zeta^2 \rangle$ & $\chi^2$ / d.o.f. \\
\hline \hline
polynomial & --- &      1.91 $\pm$ 0.22   & 9.67 $\pm$ 3.93 & 0.38 \\
\hline
power-law  & --- &      2.67 $\pm$ 0.70   & 6.84 $\pm$ 2.58 & 0.36 \\
\hline
integral   & --- &      4.4 $\pm$ 938.4   & 7.30 $\pm$ 3.14 & 0.35 \\
integral   & $a=0.5$ &  3.69 $\pm$ 1.06   & 7.14 $\pm$ 2.67 & 0.35 \\
integral   & $a=4$   &  1.92 $\pm$ 0.42   & 6.13 $\pm$ 2.36 & 0.40 \\
\hline
cosine     & --- &      3.9 $\pm$ 262.2   & 7.30 $\pm$ 3.08 & 0.35 \\
cosine     & $a=1$ &    2.07 $\pm$ 0.45   & 6.34 $\pm$ 2.49 & 0.38 \\
\hline \hline
\end{tabular}
\caption{Results of the fits in Table~\ref{tab:fit-results} for the value $S_{u
d}$ of the number sum rule and for the average value of $\zeta^2$ defined in \protect\eqref{zeta2-moment}.}
\label{tab:globalfitcomp}
\end{table}

The results of our fits for the sum rule value $f J(\zeta=0)$ are collected in
Table~\ref{tab:globalfitcomp}, where we see again the huge uncertainties on this
value obtained with the two-parameter versions of the integral and cosine
models.

Also shown in that table are the fit results for another quantity of physical
interest, namely
\begin{align}
   \label{zeta2-moment}
   \langle \zeta^2 \rangle
   &=
   \frac{\int_0^1 \dd \zeta \; \zeta^2 \, I(\zeta, y^2)}{%
         \int_0^1 \dd \zeta \; I(\zeta, y^2)}
   =
   {}- \frac{\partial^2}{\partial \omega^2} \, \hat{A}(\omega, y^2)
   \Bigg|_{\omega = 0}
   \,,
\end{align}
which may be interpreted as the average\,\footnote{%
Strictly speaking, the interpretation as an ``average'' in the statistical sense requires $I(\zeta, y^2) \ge 0$, which is the case for the central values of
all our fits, except for the polynomial model.}
value of $\zeta^2$.  In general $\langle \zeta^2 \rangle$ depends on the distance
$y$, but this dependence is absent if we adopt the factorization hypothesis
\eqref{product-ansatz}.

We see in the table that the results for $\langle \zeta^2 \rangle$ agree within
uncertainties for all our fits, much better than the results for the sum rule
value $f J(\zeta = 0)$.  To understand this, we observe in
Equation~\eqref{zeta2-moment} that $\langle \zeta^2 \rangle$ depends on the
behavior of $\hat{A}(\omega, y^2)$ around $\omega = 0$, where our lattice data
is quite precise.  By contrast, the Mellin moment at $\zeta = 0$ corresponds an
integral over all $\omega$  and is sensitive to $\hat{A}$ at large $\omega$,
where the data becomes increasingly noisy or is missing altogether.

The grand average of $\langle \zeta^2 \rangle$ over all fits is around $0.07$
(with the largest deviation from the polynomial model) and thus quite small.
This is consistent with our argument from Section~\ref{sec:moment-properties}
that $I(\zeta, y^2)$ should decrease rather quickly for $\zeta \to 1$ and makes
this statement more quantitative.

%%%%%%%%%%%%%%%%%%%%%%%%%%%%%%%%%%%%%%%%

\subsection{Testing factorization in
   \texorpdfstring{$\zeta$ and $y$}{zeta and y}}
\label{sec:local-fits}

To test whether the hypothesis \eqref{product-ansatz} of a factorized
dependence of $I(\zeta, y^2)$ on $\zeta$ and $y$ is adequate, we perform
``local'' fits of the $\zeta$ dependence in narrow bins of $y$.  Specifically,
we group the lattice data into intervals $(i - 0.5) \, a \leq |y| \leq (i + 0.5)
\, a$ with $i = 4$ to $14$ and fit them to the same models for the $\zeta$
dependence as before.

\begin{figure}
\begin{center}
\subfloat[polynomial model]{
\includegraphics[width=0.49\textwidth]{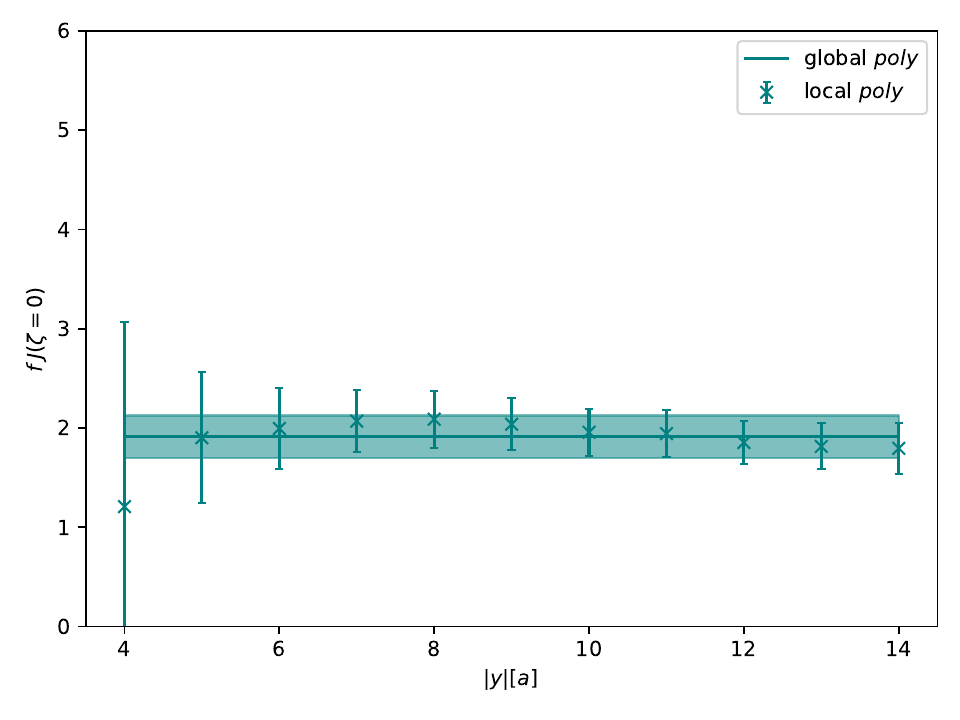}
}
\subfloat[power-law model]{
\includegraphics[width=0.49\textwidth]{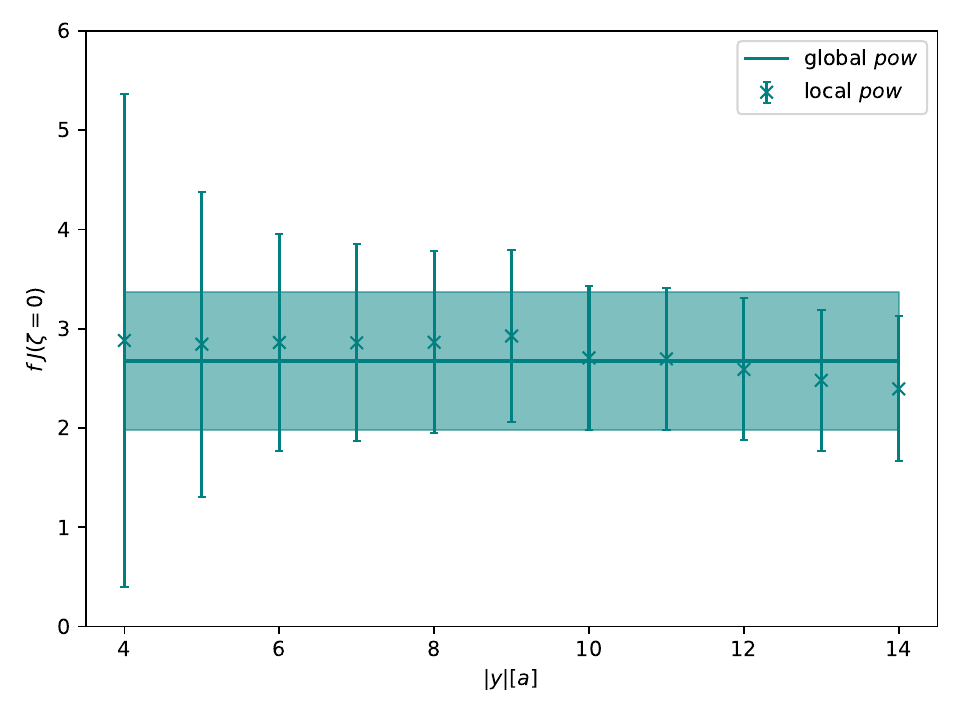}
}
\\[1em]
\subfloat[integral model with fixed $a = 4$]{
\includegraphics[width=0.49\textwidth]{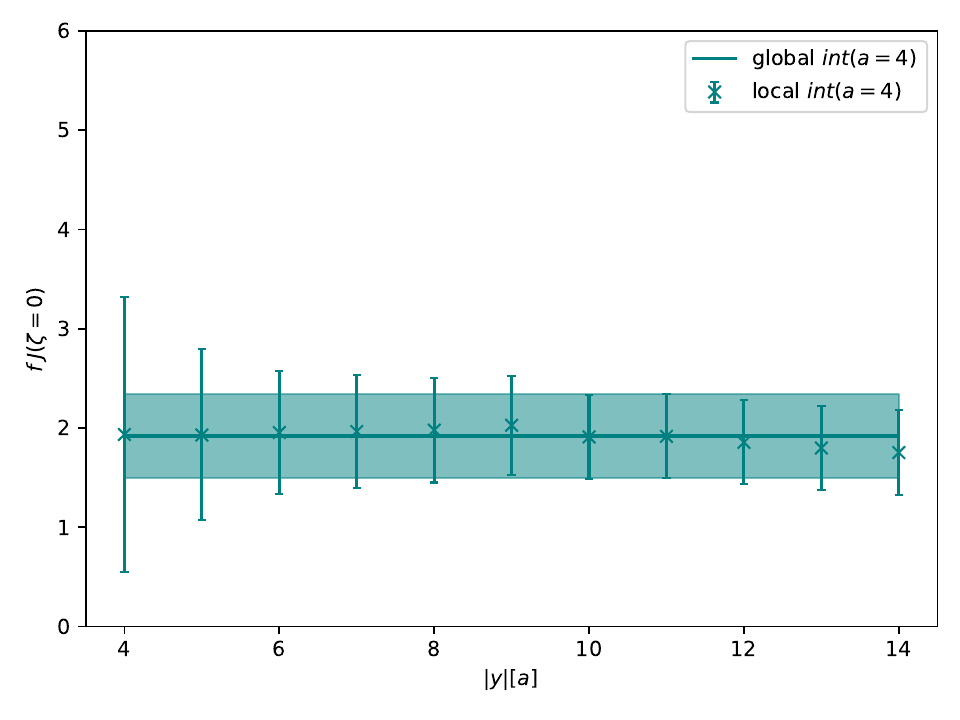}
}
\subfloat[cosine model with fixed $a = 1$]{
\includegraphics[width=0.49\textwidth]{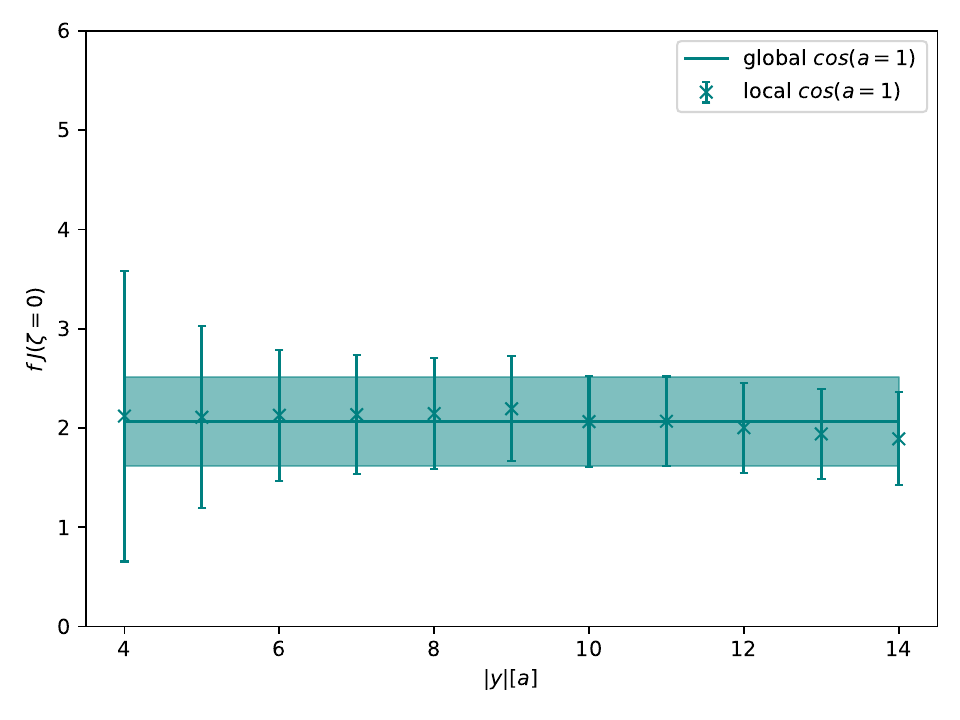}
}
\caption{\label{fig:Izeta0_model_comp} Results for $S_{u d } = f J_{u d}(\zeta =
0)$ from fits restricted to bins in $y$ of width $a$.  The bands indicate the
results of the global fits over all $y$ from $4 a$ to $14 a$.}
\end{center}
\end{figure}

\begin{figure}
\begin{center}
\subfloat[polynomial model]{
\includegraphics[width=0.49\textwidth]{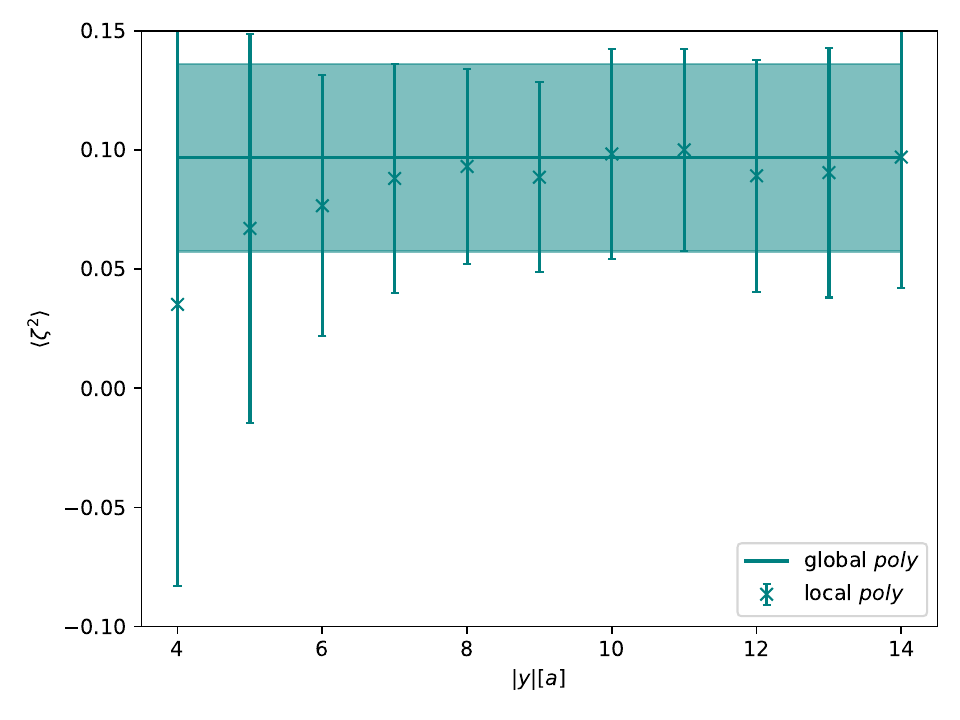}
}
\subfloat[power-law model]{
\includegraphics[width=0.49\textwidth]{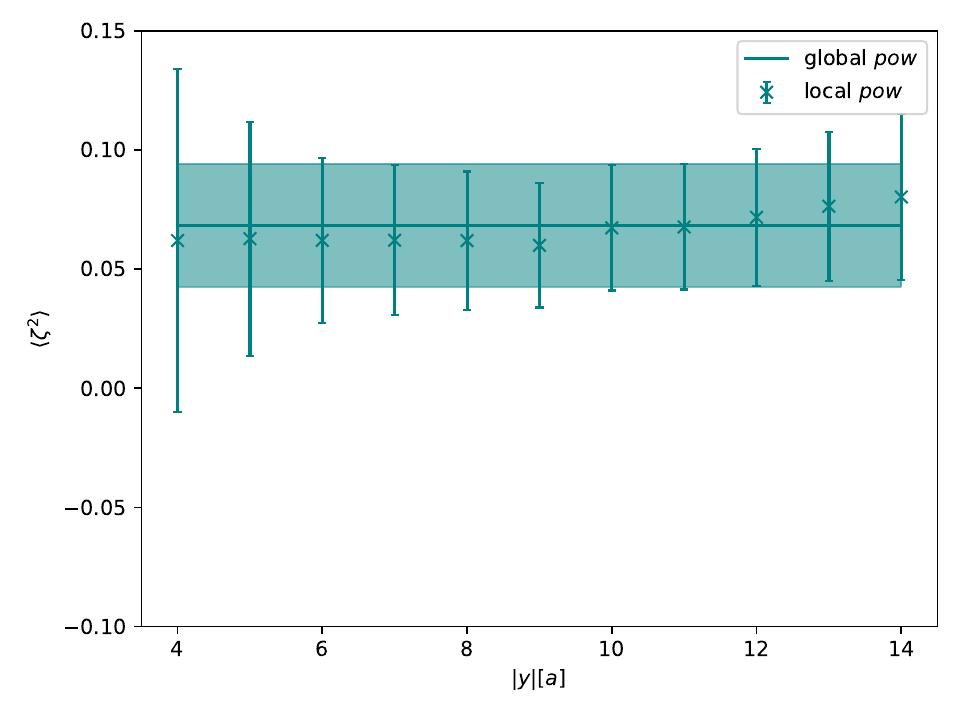}
}
\\[1em]
\subfloat[integral model with fixed $a = 4$]{
\includegraphics[width=0.49\textwidth]{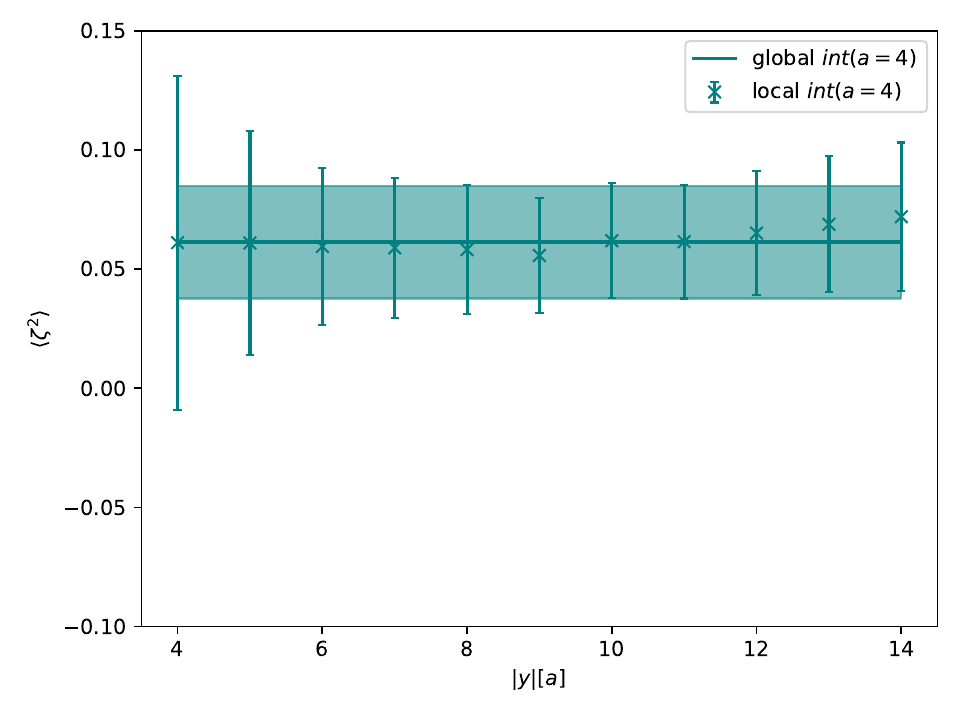}
}
\subfloat[cosine model with fixed $a = 1$]{
\includegraphics[width=0.49\textwidth]{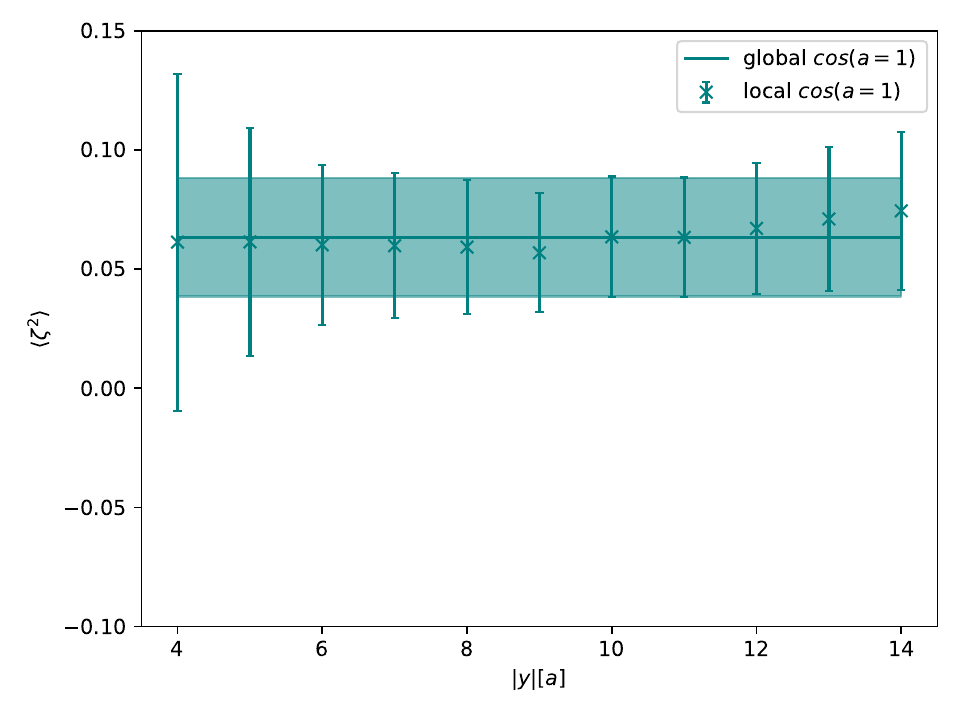}
}
\caption{\label{fig:zetasq_model_comp} As
Figure~\protect\ref{fig:Izeta0_model_comp}, but for $\langle \zeta^2 \rangle$.}
\end{center}
\end{figure}

The resulting values and errors for $f J(\zeta = 0)$ and $\langle \zeta^2
\rangle$ are shown respectively in Figures~\ref{fig:Izeta0_model_comp} and
\ref{fig:zetasq_model_comp} for those fits where the value of $f J(\zeta = 0)$
has an acceptable uncertainty and is consistent with the number sum rule.

%%%%%%%%%%%%%%%%%%%%

We see that there is no significant indication for a $y$ dependence of either
quantity.  The errors of the local fits tend to increase when $y$ becomes small,
which is due to the decreasing range of $\omega$ in that limit.  We conclude
that our factorized ansatz for the $\zeta$ and $y$ dependence is good enough
for data of the quality we have analyzed.  Note that this is equivalent to
a factorized form of the $\omega$ and $y$ dependence of the invariant
function~$A(\omega, y^2)$.

%%%%%%%%%%%%%%%%%%%%%%%%%%%%%%%%%%%%%%%%%%%%%%%%%%%%%%%%%%%%%%%%%%%%%%%%%%%%%%%%

\section{Summary}
\label{sec:summary}

Reconstructing Mellin moments of DPDs from lattice data requires the evaluation
of an integral of the data over a Ioffe time parameter $\omega$.  We can generalize
DPDs to asymmetric parton kinematics, such that their Mellin moments are
obtained by a Fourier transform of the data from $\omega$ to the skewness
parameter $\zeta$, with $\zeta = 0$ corresponding to the DPDs relevant for
double parton scattering. We have derived two important properties of the
dependence on this parameter: the Mellin moments should vanish rather quickly as
$|\zeta| \to 1$, and they can be nonanalytic functions of $\zeta$ at $0$ and
$\pm 1$.

In our previous work \cite{Bali:2021gel} we had assumed that the Mellin moments
are polynomials in $\zeta$, which is at odds with both properties just
mentioned. In the present work we have explored several model functions for the
$\zeta$ dependence that are consistent in this respect.  We fitted the
parameters of these functions to our previously generated lattice data for
unpolarized parton pairs in the flavor combination $u d$, which has the highest
statistical precision among all channels analyzed in
\cite{Bali:2021gel,Reitinger:2024ulw}.

We obtain good fits for all considered models.  With the flexible forms of our
new functions, the fits with two free parameters lead to a huge uncertainty in
the reconstructed Mellin moments at $\zeta=0$, with errors of several hundred
units for a value that should be equal to $2$ according to the number sum rule
for $u$ and $d$ quarks in the proton.  Fixing one parameter in these models such
that the sum rule is satisfied, we still obtain a good description of the data,
with a smooth behavior of the Mellin moments around $\zeta = 0$.  The
one-parameter model function $\propto (1 - |\zeta|)^a$ admits a good fit to the
data, with a Mellin moment that has a spike at $\zeta=0$ and a value just about
consistent with the sum rule. We must conclude that with the lattice data
presently at our disposal, we cannot reconstruct Mellin moments at $\zeta=0$
without additional theory input. To improve on this situation, one would need
data with significantly better precision at high $|\omega|$, i.e.\ at large
proton momenta and large distances~$y$.

Whilst one needs $\zeta=0$ to make direct contact with double parton scattering,
our results for Mellin moments at nonzero $\zeta$ still contain relevant
information about the joint distribution of two partons inside a proton.  As an
example, we find that all our fits are consistent with an average value $\langle
\zeta^2 \rangle \sim 0.07$ in the Mellin moment $I_{u d}(\zeta, y^2)$, which
indicates that parton configurations with small $|\zeta|$ are more likely than
those with large $|\zeta|$.

We recall that our studies \cite{Bali:2021gel,Reitinger:2024ulw} revealed
interesting patterns of small or large correlations between the two partons in
their spin and flavor.  These correlations were best seen in the precise
correlation functions at $\omega=0$, which according to \eqref{inv-fct-mellin}
correspond to integrating the associated Mellin moment over all $\zeta$.  With
the above finding, it is not implausible to assume that features of moments
integrated over $\zeta$ are not too different from those of the moments at
$\zeta=0$.

In the introduction we mentioned the prospect to obtain more detailed and direct
information about DPDs from lattice simulations of the associated
quasi-distributions \cite{Jaarsma:2023woo,Zhang:2023wea} (one may also think of
studies using pseudo-distributions).  Such distributions depend on three
variables that have the nature of Ioffe times, see e.g.\ Equation~(4.1) in
\cite{Jaarsma:2023woo}.  Two linear combinations of those are related by a
Fourier transform to the parton momentum fractions $x_1$ and $x_2$ in the DPD,
whilst a third combination plays a similar role as the parameter $\omega$ in the
present work.  One may therefore expect that the lessons of our study bear some
relevance for lattice investigations of DPDs beyond their Mellin moments.

%%%%%%%%%%%%%%%%%%%%%%%%%%%%%%%%%%%%%%%%%%%%%%%%%%%%%%%%%%%%%%%%%%%%%%%%%%%%%%%%

\appendix

\section{Skewed DPDs in the short-distance limit}
\label{app:splitting}

In this appendix we derive the results \eqref{splitting-lowest-moment} and
\eqref{splitting-higher-moment} used in Section~\ref{sec:moment-properties}.  As
in that section, we take $\zeta \ge 0$ without loss of generality.
The splitting process $g\to q \bar{q}$ contributes to the distribution $F_{q
\bar{q}}$ at leading order in $\alpha_s$.  Evaluation of the graph in
Figure~\ref{split-lo} gives
\begin{align}
   \label{Fqqbar-split}
   F_{q \bar{q}}^{\text{sp,LO}}(x_1, x_2, \zeta, y^2)
   &=
   C \ms f_g(x_+) \, \frac{x_+^2 + x_-^2 - \zeta^2}{2 x_+^3}
\end{align}
in the region $x_1 \ge \half\zeta$, $x_2 \ge \half\zeta$, $x_1 + x_2 \le 1$,
where we have abbreviated
\begin{align}
   C
   &=
   \frac{1}{\pi |y^2|} \, \frac{\alpha_s}{2 \pi} \, T_F
   \,,
   &
   x_{\pm}
   &=
   x_1 \pm x_2
   \,.
\end{align}
A detailed derivation of \eqref{Fqqbar-split} for the case $\zeta = 0$ is given
in sections 5.2.2 and 5.3.1 of \cite{Diehl:2011yj}, and it is easy to generalize
the corresponding calculation to nonzero $\zeta$.

Integrals of \eqref{Fqqbar-split} over the parton momentum fractions are readily
evaluated after changing variables from $x_1$ and $x_2$ to $x_+$ and $x_-$.  The
region $x_1 \ge \half\zeta$, $x_2 \ge \half\zeta$, $x_1 + x_2 \le 1$ corresponds
to $\zeta \le x_+ \le 1$, $- (x_+ - \zeta) \le x_- \le x_+ - \zeta$. Carrying
out the integral over $x_-$ and renaming $x_+$ to $z$, we obtain
\begin{align}
   \label{split-first-mom}
   \int_{\zeta/2}^{1} \dd x_1
   \int_{\zeta/2}^{1 - x_1} \dd x_2 \;
   F_{q \bar{q}}^{\text{sp,LO}}(x_1, x_2, \zeta, y^2)
   &=
   C \int_{\zeta}^1 \dd z \, f_g(z) \;
   \frac{1}{3} \biggl( 1 - \frac{\zeta}{z} \biggr)^2 \,
   \biggl( 2 +  \frac{\zeta}{z} \biggr)
\end{align}
and
\begin{align}
   \label{split-second-mom}
   &
   \int_{\zeta/2}^{1} \dd x_1 \, x_1
   \int_{\zeta/2}^{1 - x_1} \dd x_2 \, x_2 \;
   F_{q \bar{q}}^{\text{sp,LO}}(x_1, x_2, \zeta, y^2)
   \notag \\
   & \qquad
   =
   C \int_{\zeta}^1 \dd z \, z^2 \ms f_g(z) \;
   \frac{1}{60} \biggl( 1 - \frac{\zeta}{z} \biggr)^2 \,
   \biggl( 6 +  \frac{12 \zeta}{z} - \frac{2 \zeta^2}{z^2}
      -  \frac{\zeta^3}{z^3} \biggr) \,.
\end{align}
To evaluate Mellin moments of $F_{q q}$, we need this function in its full
support region in $x_1$ and $x_2$.  To this end, we use
\begin{align}
   F_{q q}^{\text{sp,LO}}(x_1, x_2, \zeta, y^2)
   &=
   \begin{cases}
      - F_{q \bar{q}}^{\text{sp,LO}}(x_1, - x_2, \zeta, y^2) &
      \text{for $x_1 \ge \half\zeta$,
                $x_2 \le -\half\zeta$, $x_1 + |x_2| \le 1$,}
      \\
      - F_{q \bar{q}}^{\text{sp,LO}}(x_2, - x_1, \zeta, y^2) &
      \text{for $x_1 \le -\half\zeta$,
                $x_2 \ge \half\zeta$, $|x_1| + x_2 \le 1$,}
      \\
      0 & \text{otherwise,}
   \end{cases}
\end{align}
where the first line follows from the relation between quark and antiquark
operators in the definition of unpolarized DPDs, the second line uses the
general symmetry relation $F_{q q}(x_1, x_2, \zeta, y^2) = F_{q q}(x_2, x_1,
\zeta, y^2)$, and the last line reflects the support properties of the
leading-order splitting graphs (see Section \ref{sec:small-y-limit}).

We thus find that the lowest Mellin moment of $F_{q q}^{\text{sp,LO}}$ is given
by $(-2)$ times the expression in \eqref{split-first-mom}.  Correspondingly, the
Mellin moment of $F_{q q}^{\text{sp,LO}}$ with weight factor $x_1 x_2$ is
given by $(+2)$ times the expression in \eqref{split-second-mom}.

%%%%%%%%%%%%%%%%%%%%%%%%%%%%%%%%%%%%%%%%%%%%%%%%%%%%%%%%%%%%%%%%%%%%%%%%%%%%%%%%

\acknowledgments

We acknowledge the CLS effort for generating the $n_f = 2 + 1$ ensembles in
\cite{Bruno:2014jqa}, one of which was used for this work.  The Feynman graphs
in this manuscript were produced with JaxoDraw \cite{Binosi:2003yf,
Binosi:2008ig}.

This work is in part supported by the Deutsche Forschungsgemeinschaft (DFG,
German Research Foundation) -- grant number 409651613 (Research Unit FOR 2926)
and grant number 493441321.

The work of Christian Zimmermann was supported in part by the U.S. Department of Energy, Office of Science, Office of Nuclear Physics, under Contract No.~DE-AC02-05CH11231 that is used to operate Lawrence Berkeley National Laboratory. This work was supported by the Alexander von Humboldt Foundation.

%%%%%%%%%%%%%%%%%%%%%%%%%%%%%%%%%%%%%%%%%%%%%%%%%%%%%%%%%%%%%%%%%%%%%%%%%%%%%%%%

\phantomsection
\addcontentsline{toc}{section}{References}

\bibliographystyle{JHEP-mod}
\bibliography{main}

\providecommand{\href}[2]{#2}\begingroup\raggedright\begin{thebibliography}{10}

\bibitem{Bartalini:2018qje}
P.~Bartalini and J.~R. Gaunt, \emph{{Multiple Parton Interactions at the LHC}},
  \href{https://doi.org/10.1142/10646}{\emph{{Adv. Ser. Direct. High Energy
  Phys.}} {\bfseries 29} (2019) }.

\bibitem{Lin:2017snn}
H.-W. Lin et~al., \emph{{Parton distributions and lattice QCD calculations: a
  community white paper}},
  \href{https://doi.org/10.1016/j.ppnp.2018.01.007}{\emph{Prog. Part. Nucl.
  Phys.} {\bfseries 100} (2018) 107}
  [\href{https://arxiv.org/abs/1711.07916}{{\ttfamily arXiv:1711.07916}}].

\bibitem{Constantinou:2020hdm}
M.~Constantinou et~al., \emph{{Parton distributions and lattice-QCD
  calculations: Toward 3D structure}},
  \href{https://doi.org/10.1016/j.ppnp.2021.103908}{\emph{Prog. Part. Nucl.
  Phys.} {\bfseries 121} (2021) 103908}
  [\href{https://arxiv.org/abs/2006.08636}{{\ttfamily arXiv:2006.08636}}].

\bibitem{Diehl:2011yj}
M.~Diehl, D.~Ostermeier and A.~Sch{\"a}fer, \emph{{Elements of a theory for
  multiparton interactions in QCD}},
  \href{https://doi.org/10.1007/JHEP03(2012)089}{\emph{JHEP} {\bfseries 03}
  (2012) 089} [\href{https://arxiv.org/abs/1111.0910}{{\ttfamily
  arXiv:1111.0910}}].
\newblock [Erratum: JHEP 03, 001 (2016)].

\bibitem{Bali:2020mij}
G.~S. Bali, L.~Castagnini, M.~Diehl, J.~R. Gaunt, B.~Gl\"a\ss{}le, A.~Sch\"afer
  et~al., \emph{{Double parton distributions in the pion from lattice QCD}},
  \href{https://doi.org/10.1007/JHEP02(2021)067}{\emph{JHEP} {\bfseries 02}
  (2021) 067} [\href{https://arxiv.org/abs/2006.14826}{{\ttfamily
  arXiv:2006.14826}}].

\bibitem{Bali:2021gel}
G.~S. Bali, M.~Diehl, B.~Gl\"a\ss{}le, A.~Sch\"afer and C.~Zimmermann,
  \emph{{Double parton distributions in the nucleon from lattice QCD}},
  \href{https://doi.org/10.1007/JHEP09(2021)106}{\emph{JHEP} {\bfseries 09}
  (2021) 106} [\href{https://arxiv.org/abs/2106.03451}{{\ttfamily
  arXiv:2106.03451}}].

\bibitem{Reitinger:2024ulw}
D.~Reitinger, C.~Zimmermann, M.~Diehl and A.~Sch{\"a}fer, \emph{{Double parton
  distributions with flavor interference from lattice QCD}},
  \href{https://doi.org/10.1007/JHEP04(2024)087}{\emph{JHEP} {\bfseries 04}
  (2024) 087} [\href{https://arxiv.org/abs/2401.14855}{{\ttfamily
  arXiv:2401.14855}}].

\bibitem{Ji:2013dva}
X.~Ji, \emph{{Parton Physics on a Euclidean Lattice}},
  \href{https://doi.org/10.1103/PhysRevLett.110.262002}{\emph{Phys. Rev. Lett.}
  {\bfseries 110} (2013) 262002}
  [\href{https://arxiv.org/abs/1305.1539}{{\ttfamily arXiv:1305.1539}}].

\bibitem{Ji:2014gla}
X.~Ji, \emph{{Parton Physics from Large-Momentum Effective Field Theory}},
  \href{https://doi.org/10.1007/s11433-014-5492-3}{\emph{Sci. China Phys. Mech.
  Astron.} {\bfseries 57} (2014) 1407}
  [\href{https://arxiv.org/abs/1404.6680}{{\ttfamily arXiv:1404.6680}}].

\bibitem{Radyushkin:2016hsy}
A.~Radyushkin, \emph{{Nonperturbative Evolution of Parton
  Quasi-Distributions}},
  \href{https://doi.org/10.1016/j.physletb.2017.02.019}{\emph{Phys. Lett. B}
  {\bfseries 767} (2017) 314}
  [\href{https://arxiv.org/abs/1612.05170}{{\ttfamily arXiv:1612.05170}}].

\bibitem{Radyushkin:2017cyf}
A.~V. Radyushkin, \emph{{Quasi-parton distribution functions, momentum
  distributions, and pseudo-parton distribution functions}},
  \href{https://doi.org/10.1103/PhysRevD.96.034025}{\emph{Phys. Rev. D}
  {\bfseries 96} (2017) 034025}
  [\href{https://arxiv.org/abs/1705.01488}{{\ttfamily arXiv:1705.01488}}].

\bibitem{Chen:2025cxr}
J.-W. Chen et~al., \emph{{Large-momentum effective theory{\textquoteright}s
  asymptotic extrapolation vs the inverse problem}},
  \href{https://doi.org/10.1103/fflw-qpcc}{\emph{Phys. Rev. D} {\bfseries 113}
  (2026) 014509} [\href{https://arxiv.org/abs/2505.14619}{{\ttfamily
  arXiv:2505.14619}}].

\bibitem{Xiong:2025obq}
A.-S. Xiong, J.~Hua, Y.-F. Ling, T.~Wei, F.-S. Yu, Q.-A. Zhang et~al.,
  \emph{{Ill-posedness in limited discrete Fourier inversion and regularization
  for quasi distributions in LaMET}},
  \href{https://doi.org/10.1140/epjc/s10052-025-15130-9}{\emph{Eur. Phys. J. C}
  {\bfseries 85} (2025) 1409}
  [\href{https://arxiv.org/abs/2506.16689}{{\ttfamily arXiv:2506.16689}}].

\bibitem{Dutrieux:2025jed}
H.~Dutrieux, J.~Karpie, C.~J. Monahan, K.~Orginos, A.~Radyushkin, D.~Richards
  et~al., \emph{{Inverse problem in the large momentum effective theory
  framework}}, \href{https://doi.org/10.1103/4f6n-vy13}{\emph{Phys. Rev. D}
  {\bfseries 113} (2026) 074524}
  [\href{https://arxiv.org/abs/2504.17706}{{\ttfamily arXiv:2504.17706}}].

\bibitem{Jaarsma:2023woo}
M.~Jaarsma, R.~Rahn and W.~J. Waalewijn, \emph{{Towards double parton
  distributions from first principles using Large Momentum Effective Theory}},
  \href{https://doi.org/10.1007/JHEP12(2023)014}{\emph{JHEP} {\bfseries 12}
  (2023) 014} [\href{https://arxiv.org/abs/2305.09716}{{\ttfamily
  arXiv:2305.09716}}].

\bibitem{Zhang:2023wea}
J.-H. Zhang, \emph{{Double Parton Distributions from Euclidean Lattice}},
  \href{https://arxiv.org/abs/2304.12481}{{\ttfamily arXiv:2304.12481}}.

\bibitem{Jaffe:1983hp}
R.~L. Jaffe, \emph{{Parton Distribution Functions for Twist Four}},
  \href{https://doi.org/10.1016/0550-3213(83)90361-9}{\emph{Nucl. Phys. B}
  {\bfseries 229} (1983) 205}.

\bibitem{Gaunt:2009re}
J.~R. Gaunt and W.~J. Stirling, \emph{{Double Parton Distributions
  Incorporating Perturbative QCD Evolution and Momentum and Quark Number Sum
  Rules}}, \href{https://doi.org/10.1007/JHEP03(2010)005}{\emph{JHEP}
  {\bfseries 03} (2010) 005} [\href{https://arxiv.org/abs/0910.4347}{{\ttfamily
  arXiv:0910.4347}}].

\bibitem{Gaunt:2012tfk}
J.~Gaunt, \emph{{Double parton scattering in proton-proton collisions}}, Ph.D.
  thesis, Cambridge U., 2012.
\newblock \href{https://doi.org/10.17863/CAM.16589}{10.17863/CAM.16589}.

\bibitem{Diehl:2018kgr}
M.~Diehl, P.~Pl{\"o}{\ss}l and A.~Sch{\"a}fer, \emph{{Proof of sum rules for
  double parton distributions in QCD}},
  \href{https://doi.org/10.1140/epjc/s10052-019-6777-5}{\emph{Eur. Phys. J. C}
  {\bfseries 79} (2019) 253}
  [\href{https://arxiv.org/abs/1811.00289}{{\ttfamily arXiv:1811.00289}}].

\bibitem{Diehl:2017kgu}
M.~Diehl, J.~R. Gaunt and K.~Sch{\"o}nwald, \emph{{Double hard scattering
  without double counting}},
  \href{https://doi.org/10.1007/JHEP06(2017)083}{\emph{JHEP} {\bfseries 06}
  (2017) 083} [\href{https://arxiv.org/abs/1702.06486}{{\ttfamily
  arXiv:1702.06486}}].

\bibitem{Bruno:2014jqa}
M.~Bruno et~al., \emph{{Simulation of QCD with N$_{f} =$ 2 $+$ 1 flavors of
  non-perturbatively improved Wilson fermions}},
  \href{https://doi.org/10.1007/JHEP02(2015)043}{\emph{JHEP} {\bfseries 02}
  (2015) 043} [\href{https://arxiv.org/abs/1411.3982}{{\ttfamily
  arXiv:1411.3982}}].

\bibitem{RQCD:2022xux}
{\scshape RQCD} collaboration, G.~S. Bali, S.~Collins, P.~Georg, D.~Jenkins,
  P.~Korcyl, A.~Sch{\"a}fer et~al., \emph{{Scale setting and the light baryon
  spectrum in N$_{f}$ = 2 + 1 QCD with Wilson fermions}},
  \href{https://doi.org/10.1007/JHEP05(2023)035}{\emph{JHEP} {\bfseries 05}
  (2023) 035} [\href{https://arxiv.org/abs/2211.03744}{{\ttfamily
  arXiv:2211.03744}}].

\bibitem{Conigli:2025qvh}
A.~Conigli, D.~Djukanovic, G.~von Hippel, S.~Kuberski, H.~B. Meyer, K.~Miura
  et~al., \emph{{Precision lattice calculation of the hadronic contribution to
  the running of the electroweak gauge couplings}},
  \href{https://arxiv.org/abs/2511.01623}{{\ttfamily arXiv:2511.01623}}.

\bibitem{Bali:2020isn}
{\scshape RQCD} collaboration, G.~S. Bali, S.~B{\"u}rger, S.~Collins,
  M.~G{\"o}ckeler, M.~Gruber, S.~Piemonte et~al., \emph{{Nonperturbative
  Renormalization in Lattice QCD with three Flavors of Clover Fermions: Using
  Periodic and Open Boundary Conditions}},
  \href{https://doi.org/10.1103/PhysRevD.103.094511}{\emph{Phys. Rev. D}
  {\bfseries 103} (2021) 094511}
  [\href{https://arxiv.org/abs/2012.06284}{{\ttfamily arXiv:2012.06284}}].
\newblock [Erratum: Phys.Rev.D 107, 039901 (2023)].

\bibitem{Bali:2016lva}
G.~S. Bali, B.~Lang, B.~U. Musch and A.~Sch{\"a}fer, \emph{{Novel quark
  smearing for hadrons with high momenta in lattice QCD}},
  \href{https://doi.org/10.1103/PhysRevD.93.094515}{\emph{Phys. Rev. D}
  {\bfseries 93} (2016) 094515}
  [\href{https://arxiv.org/abs/1602.05525}{{\ttfamily arXiv:1602.05525}}].

\bibitem{Bali:2018nde}
G.~S. Bali, P.~C. Bruns, L.~Castagnini, M.~Diehl, J.~R. Gaunt,
  B.~Gl{\"a}{\ss}le et~al., \emph{{Two-current correlations in the pion on the
  lattice}}, \href{https://doi.org/10.1007/JHEP12(2018)061}{\emph{JHEP}
  {\bfseries 12} (2018) 061}
  [\href{https://arxiv.org/abs/1807.03073}{{\ttfamily arXiv:1807.03073}}].

\bibitem{Zhang:2025hyo}
R.~Zhang, A.~V. Grebe, D.~C. Hackett, M.~L. Wagman and Y.~Zhao,
  \emph{{Kinematically enhanced interpolating operators for boosted hadrons}},
  \href{https://doi.org/10.1103/6dh4-6k4t}{\emph{Phys. Rev. D} {\bfseries 112}
  (2025) L051502} [\href{https://arxiv.org/abs/2501.00729}{{\ttfamily
  arXiv:2501.00729}}].

\bibitem{Radyushkin:1997ki}
A.~V. Radyushkin, \emph{{Nonforward parton distributions}},
  \href{https://doi.org/10.1103/PhysRevD.56.5524}{\emph{Phys. Rev. D}
  {\bfseries 56} (1997) 5524}
  [\href{https://arxiv.org/abs/hep-ph/9704207}{{\ttfamily
  arXiv:hep-ph/9704207}}].

\bibitem{Ji:1998pc}
X.-D. Ji, \emph{{Off forward parton distributions}},
  \href{https://doi.org/10.1088/0954-3899/24/7/002}{\emph{J. Phys. G}
  {\bfseries 24} (1998) 1181}
  [\href{https://arxiv.org/abs/hep-ph/9807358}{{\ttfamily
  arXiv:hep-ph/9807358}}].

\bibitem{Binosi:2003yf}
D.~Binosi and L.~Theu{\ss}l, \emph{{JaxoDraw: A Graphical user interface for
  drawing Feynman diagrams}},
  \href{https://doi.org/10.1016/j.cpc.2004.05.001}{\emph{Comput. Phys. Commun.}
  {\bfseries 161} (2004) 76}
  [\href{https://arxiv.org/abs/hep-ph/0309015}{{\ttfamily
  arXiv:hep-ph/0309015}}].

\bibitem{Binosi:2008ig}
D.~Binosi, J.~Collins, C.~Kaufhold and L.~Theu{\ss}l, \emph{{JaxoDraw: A
  Graphical user interface for drawing Feynman diagrams. Version 2.0 release
  notes}}, \href{https://doi.org/10.1016/j.cpc.2009.02.020}{\emph{Comput. Phys.
  Commun.} {\bfseries 180} (2009) 1709}
  [\href{https://arxiv.org/abs/0811.4113}{{\ttfamily arXiv:0811.4113}}].

\end{thebibliography}\endgroup
\end{document}